\pdfoutput=1
\documentclass[11pt]{article}
\usepackage[]{acl}
\usepackage{amsmath}
\usepackage{enumitem}  
\usepackage{amsfonts}
\usepackage[most]{tcolorbox}
\tcbuselibrary{breakable, skins}
\usepackage{ragged2e}
\usepackage{booktabs}
\usepackage{multirow}
\usepackage{graphicx}
\usepackage{xcolor}
\usepackage{colortbl}

\newtcolorbox{promptbox}[1]{
  breakable, enhanced,
  colback=white, colframe=black, boxrule=1.5pt, arc=3pt,
  title={\textbf{#1}}, fonttitle=\bfseries\small,
  coltitle=white, colbacktitle=black!90,
  attach boxed title to top left={yshift=-\tcboxedtitleheight/2, xshift=8pt},
  boxed title style={colback=black!90, colframe=black!90, arc=2pt, boxrule=0pt, left=4pt, right=4pt},
  top=8pt, left=8pt, right=8pt, bottom=8pt,
}
\newenvironment{pblock}{%
  \setlength{\parindent}{0pt}\setlength{\parskip}{4pt}\small\justifying}{}

\usepackage{color}
\usepackage{bbm}
\usepackage{multirow}
\usepackage{graphicx}
\usepackage{subfigure} 
\usepackage{sistyle}
\usepackage[ruled]{algorithm2e}
\usepackage{booktabs}
\usepackage{caption}
\SIthousandsep{,}
\usepackage{makecell}
\usepackage[skip=0pt]{caption}
\usepackage{amsmath}
\usepackage{hyperref}
\usepackage{array} 
\usepackage{acronym}
\usepackage[export]{adjustbox}

\usepackage{times}
\usepackage{latexsym}

\usepackage[T1]{fontenc}
\usepackage[utf8]{inputenc}

\usepackage{microtype}

\usepackage{inconsolata}

\usepackage{xcolor}
\usepackage{soul}

\newcommand{\best}[1]{\textbf{#1}}
\newcommand{\secondbest}[1]{\underline{#1}}

\newcommand{\heading}[1]{\vspace*{0.5mm}\noindent\textbf{#1.}}

\AtBeginDocument{%
  \providecommand\BibTeX{{%
    \normalfont B\kern-0.5em{\scshape i\kern-0.25em b}\kern-0.8em\TeX}}}

\makeatletter
\g@addto@macro\normalsize{%
  \abovedisplayskip 2pt plus1pt %minus1pt%
  \belowdisplayskip 2pt plus1pt
  \abovedisplayshortskip  2pt plus1pt%
  \belowdisplayshortskip  1pt plus1pt% minus1pt%
}
\makeatother

\newcommand{\benchname}{AuthorityBench\xspace}
\newcommand{\domain}{DomainAuth\xspace}
\newcommand{\entity}{EntityAuth\xspace}
\newcommand{\rag}{RAGAuth\xspace}
\newcommand{\liste}{LE\xspace}
\newcommand{\paire}{PE\xspace}

\title{AuthorityBench: Benchmarking LLM Authority Perception for Reliable Retrieval-Augmented Generation}

\author{
  \textbf{Zhihui Yao}\textsuperscript{1}\thanks{Equal contribution.}\thanks{Work done during an internship at ICT, CAS.}\quad
  \textbf{Hengran Zhang}\textsuperscript{1,2*}\quad
  \textbf{Keping Bi}\textsuperscript{1,2}\thanks{Corresponding author.}\\[4pt]
  \textsuperscript{1}State Key Laboratory of AI Safety,\\
  Institute of Computing Technology, Chinese Academy of Sciences\\
  \textsuperscript{2}University of Chinese Academy of Sciences\\[2pt]
  \texttt{1120230576@bit.edu.cn}\\
  \texttt{\{zhanghengran22z, bikeping\}@ict.ac.cn}
}

\begin{document}
\maketitle
\begin{abstract}
Retrieval-Augmented Generation (RAG) enhances Large Language Models (LLMs) with external knowledge but remains vulnerable to low-authority sources that can propagate misinformation. We investigate whether LLMs can perceive information authority—a capability extending beyond semantic understanding. To address this, we introduce \benchname, a comprehensive benchmark for evaluating LLM authority perception comprising three datasets: \domain (10K web domains with PageRank-based authority), \entity (22K entities with popularity-based authority), and \rag (120 queries with documents of varying authority for downstream evaluation). We evaluate five LLMs using three judging methods (PointJudge, PairJudge, ListJudge) across multiple output formats. Results show that ListJudge and PairJudge with PointScore output achieve the strongest correlation with ground-truth authority, while ListJudge offers optimal cost-effectiveness. Notably, incorporating webpage text consistently degrades judgment performance, suggesting authority is distinct from textual style. Downstream experiments on RAG demonstrate that authority-guided filtering largely improves answer accuracy, validating the practical importance of authority perception for reliable knowledge retrieval.\footnote{Our code and benchmark can be found at \url{https://github.com/Trustworthy-Information-Access/AuthorityBench}}
% Key findings: LLM judgments correlate strongly with ground truth; Pair/List with absolute scores outperform direct rankings; text helps PointJudge but often distracts Pair/List. On \rag, authority-guided filtering, under a unified listwise-input, pointwise-scoring pipeline, significantly improves answer accuracy and reduces misinformation compared to relevance- and utility-only baselines. Our results establish authority perception as a critical and actionable signal for reliable RAG.
\end{abstract}
\begin{figure*}[t!]
\centering
\includegraphics[width=0.9\textwidth]{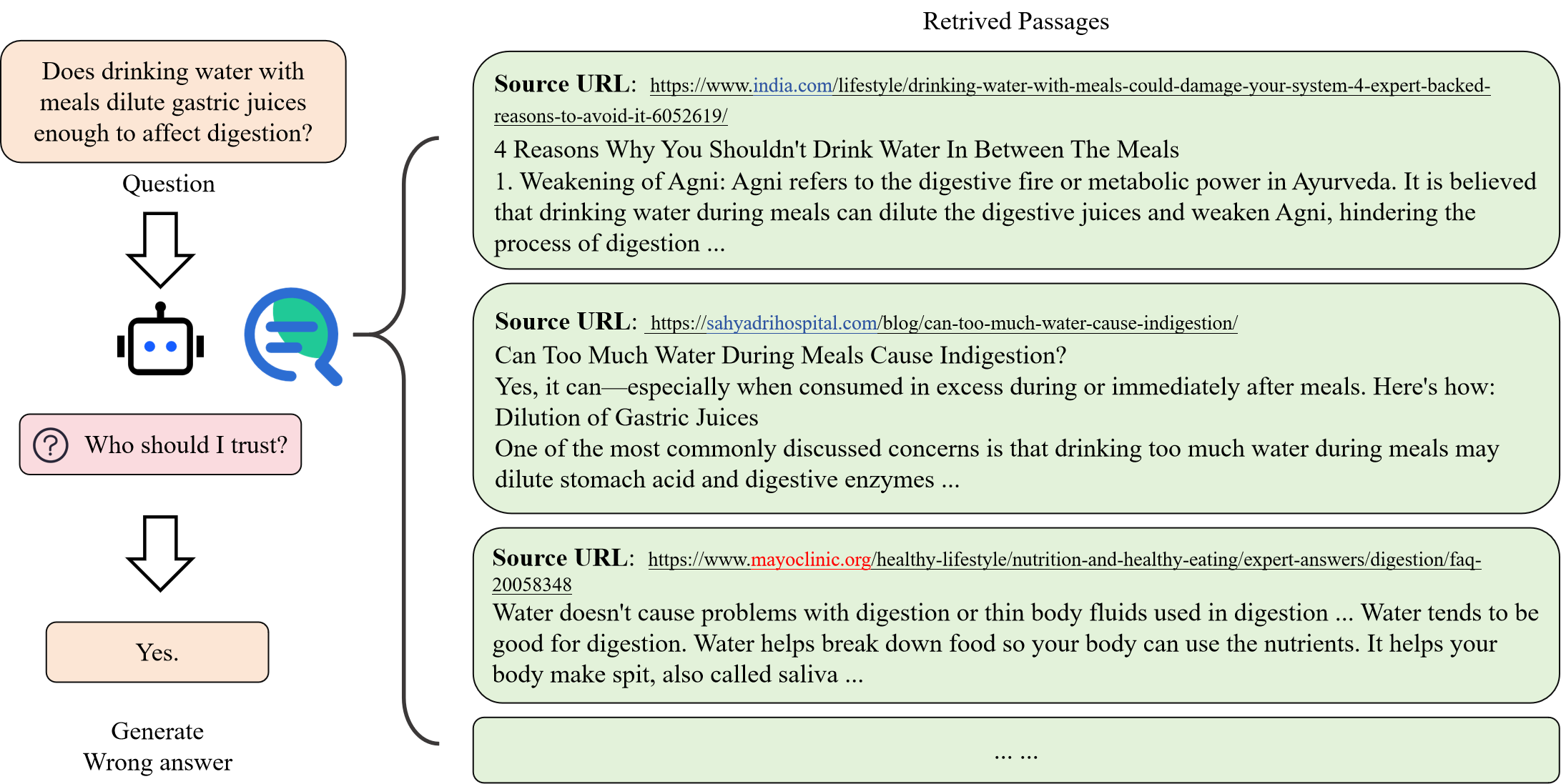}
\caption{An illustration of the authority perception challenge in RAG. When faced with conflicting information from sources of varying authority (e.g., a high-authority medical institution like Mayo Clinic vs. lower-authority lifestyle blogs), an LLM must correctly discern which source to trust to provide a reliable answer. Our work investigates this capability.}
\label{fig:show_authority_perception}
\end{figure*}

\section{Introduction}
% Large Language Models (LLMs) \cite{yang2025qwen3, team2023gemini, achiam2023gpt} have demonstrated remarkable performance across a multitude of natural language processing tasks. 
Retrieval-Augmented Generation (RAG)~\cite{lewis2020retrieval} has emerged as a dominant and effective method to mitigate outdated and hallucinations of Large Language Models (LLMs) by enabling models to
incorporate real-time, domain-specific information from external knowledge bases. 
% Many studies \cite{} point out that LLMs are highly susceptible to external knowledge during generation. If the retrieved documents contain factual inaccuracies or originate from sources with low authority, the model's output can be impacted. Therefore, it is essential to provide retrieval augmentation with high-authority documents. 
% RAG fundamentally reshapes the role of retrieval by making the LLM its primary consumer, where documents function as supporting evidence for generation rather than direct human reading \cite{zhang2025llm}. 
% Consequently, retrieval quality should be evaluated based on its contribution to generation outcomes rather than standalone document relevance. 
% Therefore, many works are focusing on how to provide documents with high context utility rather than relying on topical relevance matching for RAG \cite{ke2024bridging, jain2025modeling, zhang2025leveraging, zhang2024iterative, zhang2025distilling}.  
In practical retrieval processes for RAG systems, the sources often comprise diverse corpora, where a source with low authority can lead the system to generate misinformation \cite{schlichtkrull2024generating}, especially for queries in critical areas like health and politics. 
For example, consider the query: ``Does drinking water with meals dilute gastric juices enough to affect digestion?'' As illustrated in Figure~\ref{fig:show_authority_perception}, a RAG system might retrieve conflicting answers: a high-authority source like Mayo Clinic states that water does not cause problems, while lower-authority lifestyle blogs claim it hinders digestion. 
If the RAG cannot discern the difference in their authority and instead favors the latter simply because its prose is more fluent or persuasive, it can produce misleading or even harmful responses. 
This underscores the crucial role of authority perception in ensuring the quality and reliability of RAG systems. 

% Currently, search engines primarily determine highly authoritative websites through methods such as maintaining whitelists. 

LLMs may have already developed a preliminary ability for authority perception. For instance, LLMs inherently associate higher authority with web domains ending in ``.gov'' in our experiments. While LLMs demonstrate strong capabilities across a range of semantically related tasks \cite{yang2025qwen3, team2023gemini}, authority is a feature that extends beyond mere semantic understanding. This discrepancy raises an important question: Can LLMs effectively perform the task of authority perception? 
This leads to our central research questions: \textbf{RQ1: How can authority be defined, and the performance of authority perception be quantified?}  Furthermore, \textbf{RQ2: how can the authority perception ability of LLMs be elicited?}

% We consider two types of authority. 
For the RQ1, we consider two types of authority: (1) \textit{source authority}: This preference reflects the perceived reliability and reputation of the information source. We operationalize source authority using PageRank  \cite{rogers2002google}, a link analysis algorithm that assigns higher importance to pages that receive links from other important pages in the web graph.
(2) \textit{Entity authority}: Assertions attributed to recognized experts, institutions, or officeholders are generally considered more trustworthy than identical statements from unknown individuals. We refer to this phenomenon as entity authority. We approximate entity authority using entity popularity, a widely adopted proxy in prior works \cite{ni2025knowledge, mallen2023not}, which measures how frequently or prominently an entity appears in large corpora and knowledge sources. 
Further, we analyze the practical value of authority perception in the RAG task. 
Therefore, we introduce a comprehensive benchmark to evaluate LLMs' authority perception, i.e., \textbf{\benchname},  containing three new datasets: \textbf{\domain}, annotated with PageRank-based authority scores over 10K web domains; \textbf{\entity}, covering 22K entities across three domains with Wikipedia sitelink counts as authority proxies; and \textbf{\rag}, a downstream dataset of 120 yes/no queries paired with retrieved documents of varying authority, designed to measure the practical impact of authority perception on RAG generation quality. 
To comprehensively evaluate LLMs' authority perception, our benchmark incorporates two distinct ground truth authority labels: fine-grained (10-level) and coarse-grained (5-level) labels. 
% The evaluation is conducted at both the list level and the pair level, enabling a multi-perspective assessment. 
% We use two different ground truth labels to comprehensively evaluate different methods: fine-level (10-level) and coarse-level (5-level) labels.  Moreover, we design different evaluation settings: a listwise evaluation setting and a pairwise evaluation setting. 

For the RQ2, we evaluate five LLMs on \benchname using three distinct judging methods according to the input format, i.e., PointJudge (Each judge with one input), PairJudge (Each judge with two inputs), and ListJudge (Each judge with the list input). 
For PairJudge and ListJudge, we further design multiple output formats—either producing a direct ranking of the list/pair input (ListRank/PairRank) or assigning an absolute authority score to each item in one output (PointScore). 
% For PairJudge and ListJudge, we design multiple output formats, i.e., directly ranking list (ListRank and PairRank) or absolute authority score for each item, and then reranking based on the score (PointScore).
To analyze the context impact, each judge is applied with/without the webpage text.  
For the \domain and \entity, our findings confirm the following observations: (1) Across all settings,  ListJudge and PairJudge with PointScore output yield the strongest correlations with ground-truth authority labels; 
However, PairJudge has the highest inference cost, while ListJudge offers the best cost–effectiveness trade-off. 
% PairJudge with ScoreRank is best overall but has the highest inference cost, while ListJudge offers the best cost–effectiveness trade-off. 
% Nonetheless, comparing different output formats, ListRank and PairRank underperforms ScoreRank across settings and LLMs, likely because it forces fine-grained, relative distinctions that sharpen orderings, whereas ScoreRank’s independent scoring reduces variance but blurs subtle differences and weakens cross-item calibration. 
(2) 
Incorporating webpage text generally degrades LLM judgment under all settings, indicating authority is not equivalent to textual style, fluency, or narrative richness.   

Further, on \textbf{\rag}, we design a controlled experiment of authority-guided filtering and compare it against relevance-based \cite{sun2023chatgpt} and utility-based \cite{zhang2024large} filtering baselines: all implemented under a unified ListJudge with PointScore output and then employ top-$k$ selection. 
Experiments demonstrate that authority-guided filtering markedly boosts final answer accuracy and mitigates misguidance from untrustworthy information, validating the practical value of authority perception in RAG. 

\section{Related Work}

\vspace{-1mm}
\subsection{LLM-as-a-Judge}

A growing body of work, often termed ``LLM-as-a-Judge'' \cite{zheng2023judging, li2025generation}, which leverages LLMs to evaluate retrieved documents. This line of research has progressed through increasingly sophisticated criteria for what constitutes a ``good'' document.

\heading{From Relevance to Utility} Initial studies adapted classic information retrieval paradigms to the LLM-as-a-Judge context, focusing on \textit{relevance}. These methods, including Pointwise, Pairwise, and Listwise approaches, confirmed that LLMs can effectively rank documents based on their semantic match to a query \cite{zhuang2024setwise, sun2023chatgpt, qin2024large, ma2023zero}. For instance, Pointwise methods score each document independently, while Pairwise methods compare two documents at a time to determine the better one. However, these studies also highlighted a key limitation: a document can be highly relevant in topic but lack the specific facts needed to formulate a correct answer, leading to the ``missing evidence'' problem. Recognizing this, subsequent research introduced the concept of \textit{utility} to capture a document's actual usefulness in generating a correct answer \cite{zhang2024large, zhang2024iterative}. This marked a shift from judging topical alignment to assessing practical value, proving to be a more effective filter that significantly improves downstream QA performance \cite{zhang2024large, dewan2025llm, ke2024bridging, jain2025modeling, zhang2025utility, zhang2024iterative}.

\heading{From Content to Source Credibility} While utility focuses on document content, a parallel thread of research argues that the credibility of the source is also a critical signal. These ``credibility-aware’’ systems aim to make LLMs sensitive to source quality. For instance, \citet{pan2024not} fine-tune models on explicit, multi-dimensional credibility labels (including source authority), while \citet{deng2025cram} and \citet{hwang2025retrieval} propose methods to dynamically down-weigh information from less reliable sources based on pre-assigned scores or cross-source inconsistencies.

\subsection{Source Authority and Credibility in NLP}
% The concept of source credibility has been operationalized in various ways across NLP, often conflating a source's intrinsic authority with the veracity of its content.

In \textbf{fake news detection and claim verification}, credibility is typically a proxy for factuality. The seminal LIAR dataset, for example, uses a speaker's history of truthful or false statements as a primary signal for assessing new claims \cite{wang2017liar}. Other work explicitly uses the authority and popularity of web domains as features to predict a claim's veracity \cite{popat2016credibility, popat2017truth}. Similarly, in \textbf{media profiling}, credibility is equated with a news outlet's reporting accuracy and lack of bias, often benchmarked against human-curated ratings from organizations like Media Bias/Fact Check \cite{nakov2024survey, schlichtkrull2024generating}.

While these lines of research are crucial, they define credibility primarily in terms of content quality. In contrast, our work isolates the concept of source authority as a prior, content-agnostic property of a publisher or entity, reflecting its structural influence or global recognition. We argue that this is a distinct and fundamental dimension of credibility. While prior work has used authority as a feature, our study provides a systematic benchmark to investigate whether LLMs can perceive this signal on its own, offering a foundational analysis for this critical aspect of model intelligence.

\section{\benchname}

\subsection{Overview and Dimensions of Authority}
\label{sec:dimensions}
% Authority is a core dimension of information quality in reliable RAG. 
% When multiple webpages are equally relevant to a query, users tend to prefer results from official or well-vetted publishers—reflecting source authority. Similarly, the perceived credibility of a claim depends on who makes it: statements from recognized experts or officeholders are typically judged more trustworthy than identical statements from unknown individuals—reflecting entity authority. Because LLMs are highly sensitive to the documents they retrieve and are conditioned on \cite{shi2023large,lewis2020retrieval,mallen2023not}, exposure to low-authority content can undermine the credibility of generated answers.

% Despite the strong performance of LLMs, it remains unclear whether LLMs can detect and appropriately weight authority signals. This motivates a systematic evaluation of authority perception: if models can recognize and calibrate source- and entity-level authority, they should better mitigate the influence of low-authority inputs and produce more reliable outputs. To this end, we introduce a benchmark that probes authority perception along two axes: source authority and entity authority. 

A systematic evaluation of authority perception first requires a clear definition of authority and a method to quantify it. In information science, authority is often distinguished from credibility: while credibility is a user's subjective assessment of trustworthiness~\cite{rieh2010credibility}, authority is a more objective, source-based property~\cite{wilson1983second}. Following \citet{wilson1983second}'s notion of \textit{cognitive authority}—the idea that people grant differential trust to sources recognized as knowing what they are talking about—we define authority along two dimensions.

\heading{Source Authority} Source authority is a prior, publisher-level measure of credibility, governance, and influence, independent of any single document’s content. It captures how much a source should be trusted by default when relevance is held constant. We operationalize this dimension with \texttt{\domain} and use link-based centrality—PageRank computed over a domain-level web graph—as the primary proxy, since a domain's position in the global hyperlink structure provides an objective, scalable, and content-agnostic signal of its structural prestige.
% We operationalize this dimension with \texttt{\domain} and use link-based centrality—PageRank computed over a domain-level web graph—as the primary proxy.

\heading{Entity Authority} 
% Entity authority is a property of the speaker or source entity—typically a person or organization—that reflects expertise, mandate, accountability, and public influence, affecting how their statements are weighted regardless of the publishing venue. 
Assertions attributed to recognized experts, institutions, or officeholders are generally considered more trustworthy than identical statements from unknown individuals.
% We operationalize this dimension with \texttt{\entity} and use the entity’s cross-lingual footprint—the number of interlanguage sitelinks on Wikipedia, i.e., entity popularity, as a practical proxy for global recognition.
We operationalize this dimension with \texttt{\entity} and use the entity's cross-lingual footprint—the number of interlanguage sitelinks on Wikipedia, i.e., entity popularity—as a practical proxy for global recognition, since this metric captures the breadth of an entity's presence across cultures and languages, serving as a strong indicator of its standing in the global knowledge graph.
% , and has been widely adopted as a proxy for entity prominence in prior work~\cite{ni2025how, mallen2023not}.

\begin{figure}[t]
\centering
\includegraphics[width=0.9\columnwidth]{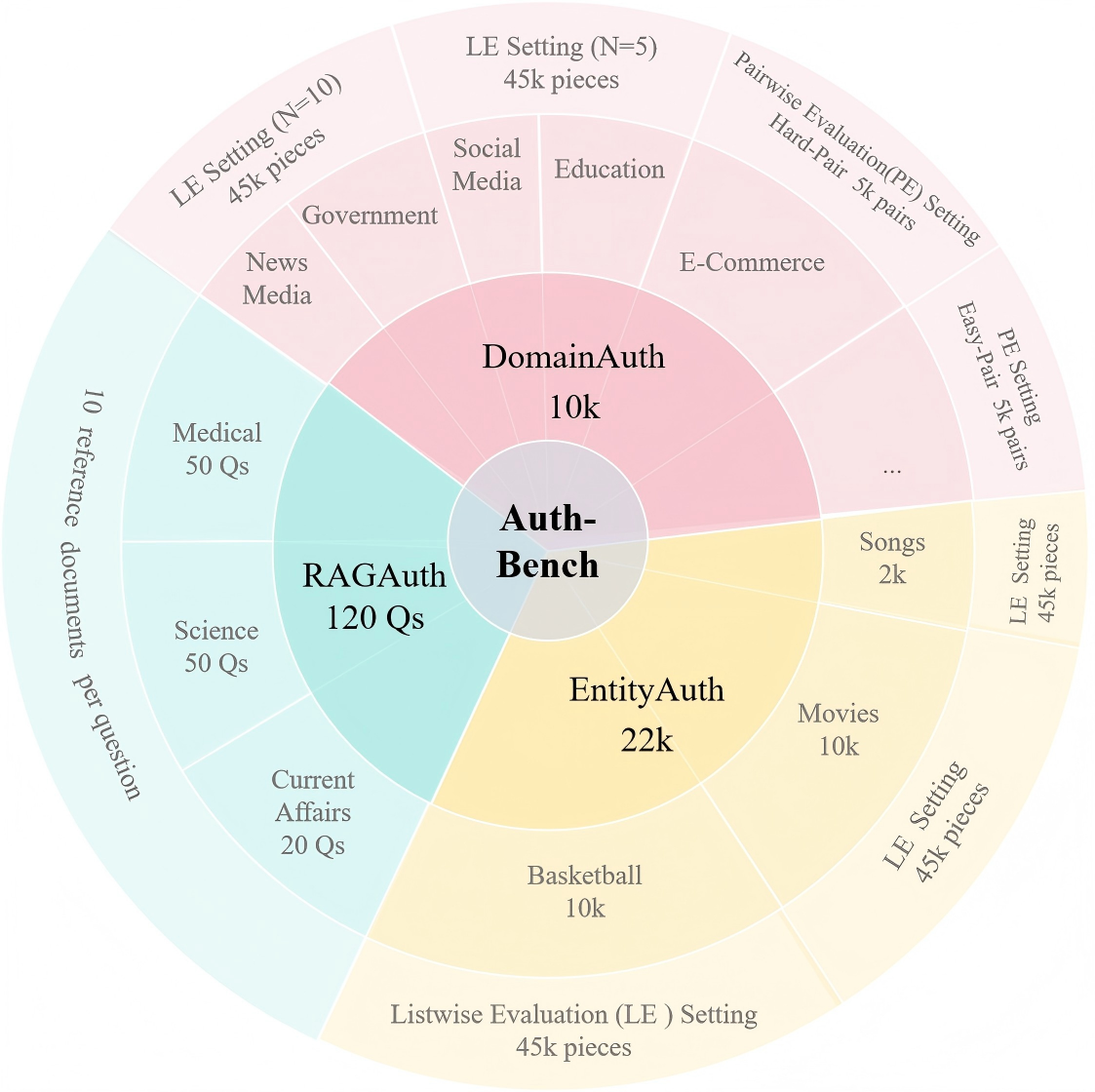}
\caption{An overview of \benchname. It consists of three sub-tasks: \domain for source authority, \entity for entity authority, and \rag for downstream RAG evaluation. The outer rings show the topic distribution within each dataset.}
\label{fig:dataset_overview}
\end{figure}

% In summary, these two dimensions offer a principled basis for evaluating LLMs’ authority perception.

% Authority in information retrieval is a multi-faceted construct: it can refer to the structural importance of an information \textit{source} within a network, or to the global recognition of a named \textit{entity} as a subject of knowledge \citep{rieh2010credibility, wilson1983second}. 

% To systematically evaluate the ability of LLMs' authority perception, we introduce \benchname, a benchmark comprising three sub-tasks. 
% Authority in information science is a multi-faceted construct: it can refer to the structural importance of an information \textit{source} within a network, or to the global recognition of a named \textit{entity} as a subject of knowledge \citep{rieh2010credibility, wilson1983second}. 
% Motivated by this distinction, we define authority along two dimensions.

% \heading{Source Authority} It measures the trustworthiness and influence of a web \textit{source} as determined by its position in the hyperlink network. 
% We instantiate this dimension in \textbf{DomainAuth}. 
% PageRank \citep{page1999pagerank} is the typical proxy of this concept. 

% \heading{Entity Authority} It measures the global recognition of a named \textit{entity} across cultures and languages, proxied by the number of Wikipedia sitelinks. 
% We instantiate this dimension in \textbf{EntityAuth}.

% \textbf{RAGAuth} further measures the practical utility of authority perception in downstream RAG tasks. Table~\ref{tab:benchmark_overview} provides an overview of all three sub-tasks.

\subsection{Benchmark Construction} 
To evaluate authority perception, we construct two sub-benchmarks—\domain and \entity—by adapting instances from existing datasets \cite{chen2019tiangong, ni2025knowledge}. These target source-level and entity-level authority, respectively. To assess the practical value of authority perception in downstream RAG, we further introduce the RAGAuth sub-benchmark. Figure~\ref{fig:dataset_overview} presents an overview of our proposed \benchname.

\heading{Data Source}
\domain is constructed from the TianGong-ST web-scale search corpus \citep{chen2019tiangong}, comprising 10,000 diverse web domains spanning news media, government, education, e-commerce, and social media. We use Google Toolbar PageRank scores obtained via public SEO tools as ground-truth authority labels. PageRank \citep{page1999pagerank} is the most recognized algorithm for quantifying web domain importance based on the hyperlink structure, where pages endorsed by authoritative sources are themselves considered authoritative, providing an objective and scalable signal for domain prestige.

\heading{Authority Annotation}
The Google Toolbar PageRank\footnote{\url{https://www.openpagerank.com}} assigns integer scores from 0 to 10. Because domains with a score of 10 constitute less than 0.01\% of the corpus, we exclude this tier and retain scores in the range 0--9, yielding 10 discrete authority levels. 
We adopt two label granularities for the final annotations: (i) a fine-grained scheme that preserves the 10 levels (0–9), and (ii) a coarse-grained scheme that collapses adjacent levels into a five-point scale (0\&1$\rightarrow$0, 2\&3$\rightarrow$1, 4\&5$\rightarrow$2, 6\&7$\rightarrow$3, 8\&9$\rightarrow$4).

\subsubsection{\entity}

\heading{Data Source}
\entity is derived from the entity-centric QA dataset introduced by \citet{ni2025knowledge}, which covers three domains: basketball players, movies, and songs. We sample 2,000 entities from the songs domain, and 10,000 each from the movies and basketball domains to form our corpus. Following \citet{ni2025knowledge}, we use the number of Wikidata sitelinks---the count of language editions of Wikipedia that have a dedicated page for the entity---as a proxy for its global prominence and recognition. This metric robustly captures an entity's cognitive breadth across cultures, serving as a strong indicator of its authority in the global knowledge graph.

\heading{Authority Annotation}
The raw sitelink counts exhibit a strong power-law distribution, similar to the hyperlink structures underlying PageRank. To create a standardized 0--9 score comparable to \domain, and inspired by the widely-held view that Google's Toolbar PageRank employs a logarithmic scale \citep{rogers2002google, page1999pagerank}, we apply logarithmic binning to the raw sitelink counts. For a domain with a count range $[s_{\min}, s_{\max}]$, we map each count $s$ to a score via:
% {\small
%     \begin{multline}
%         \text{score}(s) \;=\;
%         \texttt{np.digitize} \!( \\
%         s,\;\texttt{np.logspace}(\log_{10} s_{\min},\, \log_{10} s_{\max},\, 11)) - 1. 
%         \label{eq:log_bins}
%     \end{multline}
% }
\begin{multline}
    \text{score}(s) = \texttt{np.digitize}\bigl( s,\ \texttt{np.logspace}\bigl( \bigr. \\
    \log_{10} s_{\min},\ \log_{10} s_{\max},\ 11 \bigl) \bigr) - 1.
    \label{eq:log_bins}
\end{multline}

This process creates 10 log-equal-width intervals, assigning each entity to a bin based on its order-of-magnitude prominence rather than linear differences in raw counts.
% For instance, in the songs domain, an entity with 10 sitelinks scores 2, while one with 40 scores 8.

\subsubsection{\rag}
% \paragraph{Task Design.} While DomainAuth and EntityAuth evaluate authority perception as an intrinsic capability, RAGAuth measures its practical downstream utility: can a model leverage authority perception for more reliable RAG? 
To investigate whether a model can leverage authority perception to achieve more reliable RAG, we manually constructed the \rag dataset by the authors. 

\heading{Query Curation}
To enable accurate evaluation of answer correctness, the dataset comprises 120 yes/no questions spanning topics such as current affairs and medicine—domains susceptible to online misinformation, where low-authority sources may provide plausible but incorrect answers. 

\heading{Document Curation} 
For each question, we curated 10 web documents and their source URLs retrieved via Bing search \footnote{\url{https://www.bing.com}}. We deliberately mix high‑ and low‑PageRank sources to construct a hybrid setting, simulating the real‑world web environment.

\subsection{Evaluation Setting}
% A key principle of our benchmark is decoupling the evaluation setting (input format) from the judge (elicitation method), ensuring direct comparability.
% We employ two evaluation settings:

\subsubsection{Authority Perception Evaluation}
\heading{Listwise Evaluation (\liste) Setting} Each instance comprises $N$ inputs. Each list is constructed by sampling domains/entities to cover all authority levels, i.e., one item from each level, thereby evaluating authority ranking across multiple domains/entities from both fine-grained and coarse-grained perspectives. For LE, we use Spearman's $\rho$~\cite{spearman1961proof} and Kendall's $\tau$~\cite{kendall1938new} as rank correlation metrics.

% The evaluation metrics are Kendall’s tau \cite{kendall1938new} and Spearman's correlation coefficient \cite{spearman1904proof} .

% We use both list sizes to: (1) verify whether the relative performance of the three judgers is consistent as the number of candidates changes; (2) enable exhaustive pairwise evaluation for $N{=}5$ lists ($\binom{5}{2}=10$ pairs), yielding an exact ranking to validate the anchor-based approximation in the $N{=}10$ setting. The ground-truth ranking for each list is determined by the corresponding PageRank scores.

\heading{Pairwise Evaluation (\paire) Setting} Each instance comprises two inputs, and the task is to determine which source is more authoritative. To better assess LLMs’ authority perception, we define two difficulty regimes based on the absolute difference in authority labels on the 10-point (0–9) scale: (1) easy pairs, where the PageRank value difference exceeds 5; and (2)  hard pairs, where the PageRank value difference is less than 2. For PE, we report paired-preference accuracy as the performance metric. 
% 10,000 pairs, stratified into 5,000 \textbf{hard pairs} (0 $<$ $\Delta PR$ $\leq$ 2) and 5,000 \textbf{easy pairs} ($\Delta PR$ $\geq$ 5) for fine-grained analysis.

% For the LE setting on \domain and \entity, we use Spearman's $\rho$~\cite{spearman1904proof} and Kendall's $\tau$~\cite{kendall1938new} as rank correlation metrics. For the PE setting, we report paired-preference accuracy.

\subsubsection{RAG Evaluation} To evaluate the impact on RAG, we employ various criteria for selecting reference documents. For \rag, we adopt answer accuracy as the metric to measure the quality of generated outputs.

% For the LE setting  on \domain and \entity, we use Spearman's $\rho$~\cite{spearman1904proof} and Kendall's $\tau$~\cite{kendall1938new} as rank correlation metrics. For the PE setting, we report paired-preference accuracy. For \rag, we report answer accuracy. 

\subsection{Data Statistics}

\begin{figure}[t]
\centering
\includegraphics[width=\columnwidth]{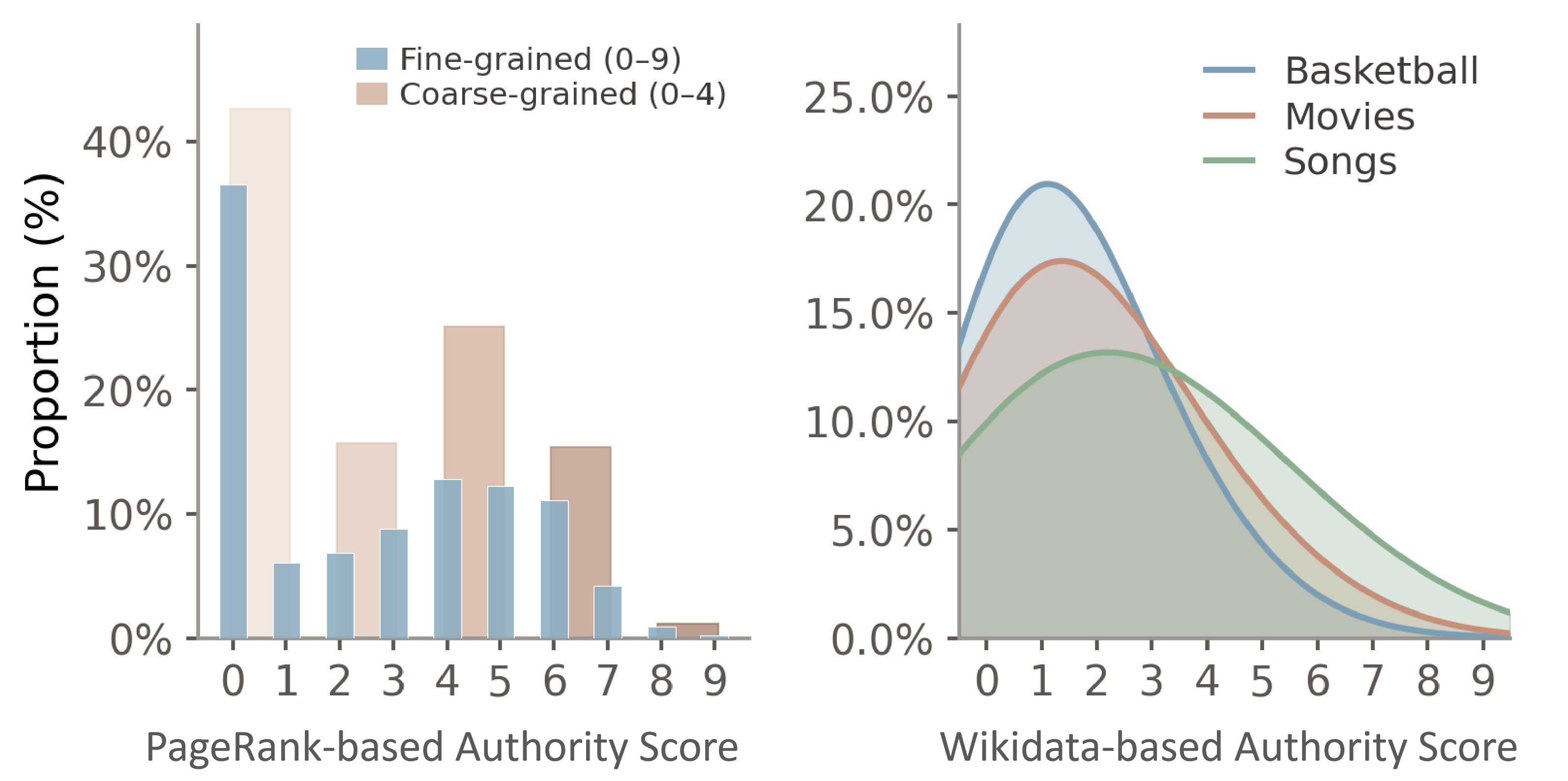}
\caption{Authority score distribution in \domain and \entity.}
\label{fig:domainauth_dist}
\end{figure}

Table~\ref{tab:data_statistics} in the Appendix summarizes the scale of all three sub-benchmarks. As shown in Figures~\ref{fig:domainauth_dist}, both \domain and \entity exhibit pronounced long-tail authority distributions: the majority of domains and entities cluster at low scores, while only a small fraction attain high authority. The coarse-grained (0--4) mapping for \domain partially alleviates this imbalance by merging adjacent levels.

\begin{table*}[t]
\centering
\renewcommand{\arraystretch}{0.8}
\resizebox{\textwidth}{!}{%
\begin{tabular}{llcccccccccc}
\toprule
\multicolumn{2}{c}{} &
  \multicolumn{2}{c}{\textbf{PointJudge}} &
  \multicolumn{4}{c}{\textbf{ListJudge}} &
  \multicolumn{4}{c}{\textbf{PairJudge}} \\
\cmidrule(lr){3-4}\cmidrule(lr){5-8}\cmidrule(lr){9-12}
\multicolumn{2}{c}{} &
  & &
  \multicolumn{2}{c}{\textbf{ListRank}} &
  \multicolumn{2}{c}{\textbf{PointScore}} &
  \multicolumn{2}{c}{\textbf{PairRank}} &
  \multicolumn{2}{c}{\textbf{PointScore}} \\
\cmidrule(lr){5-6}\cmidrule(lr){7-8}\cmidrule(lr){9-10}\cmidrule(lr){11-12}
\textbf{Model} & \textbf{Ctx} &
  $\rho$ & $\tau$ & $\rho$ & $\tau$ & $\rho$ & $\tau$ & $\rho$ & $\tau$ & $\rho$ & $\tau$ \\
\midrule
\multirow{2}{*}{Qwen3-8B}
  & w/o & 41.97 & 33.33 & 54.01 & 41.54 & \secondbest{67.11} & \secondbest{52.44} & 15.18 & 11.56 & \best{71.35} & \best{56.33} \\
  & w/  & \secondbest{45.01} & \secondbest{36.51} & 20.24 & 15.12 & 29.29 & 22.17 & 11.38 & 8.48 & \best{63.91} & \best{50.12} \\
\midrule
\multirow{2}{*}{Qwen3-14B}
  & w/o & 71.97 & 56.37 & 72.02 & 57.05 & \secondbest{73.09} & \secondbest{57.95} & 70.21 & 54.84 & \best{73.43} & \best{58.33} \\
  & w/  & \best{72.73} & \best{57.55} & 58.18 & 45.26 & 61.67 & 48.46 & 64.78 & 50.01 & \secondbest{67.99} & \secondbest{53.73} \\
\midrule
\multirow{2}{*}{Qwen3-32B}
  & w/o & 73.72 & 58.22 & 73.63 & 58.67 & \secondbest{74.41} & \secondbest{59.38} & 72.10 & 56.66 & \best{75.28} & \best{60.13} \\
  & w/  & \best{73.57} & \best{58.40} & 55.85 & 43.63 & 63.10 & 49.97 & 66.32 & 51.68 & \secondbest{69.93} & \secondbest{55.51} \\
\midrule
\multirow{2}{*}{Llama-3-8B-Instruct}
  & w/o & 63.87 & 49.56 & 57.53 & 44.09 & \best{66.08} & \best{51.12} & 61.05 & 46.48 & \secondbest{64.83} & \secondbest{49.93} \\
  & w/  & \secondbest{60.59} & \secondbest{47.00} & 34.30 & 26.05 & 27.46 & 20.73 & 49.06 & 37.13 & \best{64.97} & \best{50.08} \\
\midrule
\multirow{2}{*}{Llama-3.1-8B-Instruct}
  & w/o & \secondbest{64.78} & \secondbest{49.91} & 58.91 & 45.30 & \best{66.52} & \best{51.53} & 63.82 & 49.12 & 63.24 & 48.73 \\
  & w/  & \secondbest{57.70} & \secondbest{44.61} & 39.16 & 29.85 & 43.00 & 33.06 & 51.96 & 39.31 & \best{62.88} & \best{48.48} \\
\bottomrule
\end{tabular}
}
\caption{Results on \domain (fine-grained, 10-level) reported as Spearman's $\rho$ (\%) and Kendall's $\tau$ (\%). \textbf{Bold}: best; \underline{underline}: second best, per LLM.}
\label{tab:domain_auth_fine}
\end{table*}

\section{Experiments}
\label{sec:experiments}
% In this section, we conduct a comprehensive evaluation of various LLMs on our proposed \benchname to assess their authority perception capabilities. 
% We investigate how different judging methods and the inclusion of textual information affect model performance across both \domain and \entity.
\vspace{-1mm}
\subsection{Experimental Setup}
\heading{Evaluated LLMs} 
We evaluate five representative LLMs spanning different scales and model families: Qwen3-8B, Qwen3-14B, Qwen3-32B~\cite{yang2025qwen3}, Llama-3-8B-Instruct, and Llama-3.1-8B-Instruct~\cite{grattafiori2024llama}.
 % Additionally, we include two closed-source models, Claude-Sonnet-4-5~\cite{claude3} and Gemini-3-Flash~\cite{gemini1.5} for comparison. 
 To ensure a fair comparison, the temperature is uniformly set to 0, and the thinking function is disabled for Qwen3.  

% \heading{Metrics} For the LE setting  on \domain and \entity, we use Spearman's $\rho$~\cite{spearman1904proof} and Kendall's $\tau$~\cite{kendall1938new} as rank correlation metrics. For the PE setting, we report paired-preference accuracy. For \rag, we report answer accuracy. 

% \heading{Experimental Design}
% On \domain and \entity, we apply all three judging methods---Pointwise-Judge, Pairwise-Judge, and Listwise-Judge---under two input conditions: with and without the source's webpage text (\textit{w/ text} and \textit{w/o text}). For each judge, we compare two output formats: the native direct output format and a unified score-based format. For \domain and \entity, experiments are conducted under both fine-grained (10-level) and coarse-grained (5-level) label settings. On \rag, we evaluate authority-guided filtering against relevance-based and utility-based baselines under a unified listwise-input, pointwise-scoring pipeline.
\heading{Analyzed Method}
We adopt three LLM-based judging paradigms, corresponding to different input formats: 

\heading{(1) PointJudge} Assigns an absolute authority score to each input. When applied to listwise and pairwise evaluations, we induce rankings and preferences by ordering items according to their scores.

\heading{(2) PairJudge} Given two inputs, PairJudge decides which is more authoritative. We employ two output modes: (i) PairRank (PR), which directly returns the preferred order of the pair; and (ii) PointScore (S), which assigns an absolute authority score to each input and ranks based on authority score. 
% To derive a full listwise ranking from pairwise judgments, we adapt the strategy to the label granularity. 
For fine-grained (10-level) labels, where exhaustive pairwise comparison is computationally expensive, we use an anchor-based approximation: for each list of 10 items, we randomly select 5 anchors. Each non-anchor is compared against all 5 anchors, and each anchor is compared against the other 4 anchors. These comparisons suffice to infer the complete ranking. For coarse-grained (5-level) labels, we perform exhaustive pairwise comparisons. 
To construct the final re-ranked list from PairJudge outputs upon the listwise evaluation setting, we consider two methods: (1) \texttt{AverageScore}, which computes the mean authority score for each item and ranks by this average; and (2) \texttt{BubbleSort}, which orders items using pairwise preferences. In PairRank, BubbleSort is applied directly to the predicted pairwise orderings. In PointScore, we evaluate both \texttt{AverageScore} and \texttt{BubbleSort}. Across datasets, \texttt{AverageScore} consistently outperforms \texttt{BubbleSort} (see Tables~\ref{tab:pairjudge_domain_fine}, \ref{tab:pairjudge_domain_coarse},~\ref{tab:pairjudge_entity_fine} and~\ref{tab:pairjudge_entity_coarse} in Appendix~\ref{app:pairjudge_comparison}). The reason may be that \texttt{AverageScore} is more robust to intransitive or locally inconsistent judgments than \texttt{BubbleSort}. Therefore, we employ \texttt{AverageScore} for PointScore mode in the main experiments. 

% We attribute this to PointScore’s robustness to intransitive or locally inconsistent judgments: BubbleSort is sensitive to contradictions (e.g., A > B, B > C, but C > A), whereas averaging provides a more stable global estimate that mitigates such noise. Therefore, we employ AverageScore in the main experiments. 

% For a listwise evaluation setting, the final ranking is derived using the bubble sort method based on pairwise comparisons. Moreover, for fine-grained labels on the 10-level scale, where exhaustive comparison requires 45 pairs, we employ an anchor-based approximation: each item is compared against $K$ preselected anchors, the win numbers are computed, and items are ranked by these numbers. 

\heading{(3) ListJudge} Given a list of inputs, produces a ranking in descending order of authority. We support two output modes: (i) \textbf{ListRank (LR)}: directly ranking (outputting the re-ranked sequence), and (ii) \textbf{PointScore (S)}: the judge assigns an absolute authority score to each item, and the final ranking is obtained by sorting these scores. 

% Moreover, each judge is tested \textbf{without webpage text (w/o text)} (using only the domain name) and \textbf{with webpage text (w/ text)} (using the domain name together with a text snippet from its webpage) to analyze the influence of textual evidence. All the prompts are shown in Appendix \ref{sec:prompts}. 

Moreover, each judge is tested under two settings: (i) \textbf{without webpage text (w/o text)} (relying solely on the domain name) and (ii) \textbf{with webpage text (w/ text)} (using the domain name together with a text snippet from its corresponding webpage). This allows us to analyze the influence of textual evidence. All prompts are shown in Appendix \ref{sec:prompts}.

\begin{table*}[t]
\centering
\renewcommand{\arraystretch}{0.8}
\resizebox{\textwidth}{!}{%
\begin{tabular}{llcccccccccc}
\toprule
\multicolumn{2}{c}{} &
  \multicolumn{2}{c}{\textbf{PointJudge}} &
  \multicolumn{4}{c}{\textbf{ListJudge}} &
  \multicolumn{4}{c}{\textbf{PairJudge}} \\
\cmidrule(lr){3-4}\cmidrule(lr){5-8}\cmidrule(lr){9-12}
\multicolumn{2}{c}{} & & &
  \multicolumn{2}{c}{\textbf{ListRank}} &
  \multicolumn{2}{c}{\textbf{PointScore}} &
  \multicolumn{2}{c}{\textbf{PairRank}} &
  \multicolumn{2}{c}{\textbf{PointScore}} \\
\cmidrule(lr){5-6}\cmidrule(lr){7-8}\cmidrule(lr){9-10}\cmidrule(lr){11-12}
\textbf{Model} & \textbf{Ctx} &
  $\rho$ & $\tau$ & $\rho$ & $\tau$ & $\rho$ & $\tau$ & $\rho$ & $\tau$ & $\rho$ & $\tau$ \\
\midrule
\multirow{2}{*}{Qwen3-8B}
  & w/o & 57.29 & 49.86 & 65.60 & 55.52 & \secondbest{67.58} & \secondbest{57.47} & 13.34 & 10.69 & \best{70.90} & \best{60.22} \\
  & w/  & \secondbest{54.89} & \secondbest{47.35} & 27.81 & 22.88 & 44.73 & 37.45 & 8.73 & 7.08 & \best{64.84} & \best{54.83} \\
\midrule
\multirow{2}{*}{Qwen3-14B}
  & w/o & 67.11 & 55.82 & 71.58 & 60.87 & \secondbest{72.40} & \secondbest{61.85} & 65.65 & 54.65 & \best{73.16} & \best{62.53} \\
  & w/  & \best{68.45} & \secondbest{57.72} & 63.01 & 53.28 & 62.20 & 52.84 & \secondbest{68.30} & \best{57.77} & 66.55 & 56.64 \\
\midrule
\multirow{2}{*}{Qwen3-32B}
  & w/o & 70.40 & 59.30 & \secondbest{73.92} & \secondbest{63.35} & 73.49 & 63.04 & 72.88 & 62.10 & \best{75.52} & \best{65.23} \\
  & w/  & \best{70.40} & \best{59.87} & 59.74 & 50.71 & 63.31 & 54.40 & \secondbest{69.82} & \secondbest{59.38} & 66.55 & 56.64 \\
\midrule
\multirow{2}{*}{Llama-3-8B-Instruct}
  & w/o & \secondbest{62.54} & \secondbest{51.99} & 55.34 & 45.56 & 61.26 & 50.98 & 55.09 & 44.64 & \best{64.30} & \best{53.66} \\
  & w/  & \best{58.76} & \best{49.15} & 40.84 & 33.52 & 39.41 & 32.59 & 52.04 & 42.64 & \secondbest{56.07} & \secondbest{47.17} \\
\midrule
\multirow{2}{*}{Llama-3.1-8B-Instruct}
  & w/o & 62.73 & 51.99 & 53.30 & 43.84 & 63.52 & 52.96 & \best{64.26} & \secondbest{53.29} & \secondbest{63.84} & \best{53.32} \\
  & w/  & 56.58 & \secondbest{46.92} & 47.02 & 38.69 & 45.24 & 37.87 & \secondbest{57.08} & 46.91 & \best{58.69} & \best{49.45} \\
\bottomrule
\end{tabular}
}
\caption{Results on \domain (coarse-grained, 5-level) reported as Spearman's $\rho$ (\%) and Kendall's $\tau$ (\%). \textbf{Bold}: best; \underline{underline}: second best, per model.}
\label{tab:domain_auth_coarse}
\end{table*}

\begin{table}[t]
\centering
\small
\resizebox{\columnwidth}{!}{%
\begin{tabular}{llcccccc}
\toprule
\multicolumn{2}{c}{\multirow{2}{*}{\textbf{Model}}} &
  \multicolumn{2}{c}{\textbf{PairJudge(PR)}} &
  \multicolumn{2}{c}{\textbf{PairJudge(S)}} &
  \multicolumn{2}{c}{\textbf{PointJudge}} \\
\cmidrule(lr){3-4}\cmidrule(lr){5-6}\cmidrule(lr){7-8}
\multicolumn{2}{c}{} &
  \textbf{w/o ctx} & \textbf{w/ ctx} &
  \textbf{w/o ctx} & \textbf{w/ ctx} &
  \textbf{w/o ctx} & \textbf{w/ ctx} \\
\midrule
\multirow{2}{*}{Qwen3-8B}
  & Easy & 44.28 & 21.40 & \best{83.92} & \secondbest{67.34} & \secondbest{56.88} & \best{77.02} \\
  & Hard & \secondbest{14.96} & 16.20 & 14.46 & \secondbest{18.72} & \best{31.36} & \best{54.18} \\
\midrule
\multirow{2}{*}{Qwen3-14B}
  & Easy & \best{99.76} & \secondbest{95.62} & 97.02 & 91.94 & \secondbest{98.52} & \best{99.38} \\
  & Hard & \best{51.94} & \best{84.18} & \secondbest{40.24} & \secondbest{76.64} & 39.68 & 75.50 \\
\midrule
\multirow{2}{*}{Qwen3-32B}
  & Easy & \best{99.42} & \secondbest{95.20} & \secondbest{98.68} & 94.74 & 98.54 & \best{99.54} \\
  & Hard & \best{37.32} & \secondbest{72.98} & 35.64 & \best{75.12} & \secondbest{36.38} & 72.84 \\
\midrule
\multirow{2}{*}{Llama-3-8B}
  & Easy & \best{98.72} & 81.52 & \secondbest{96.32} & \best{90.98} & 92.76 & \secondbest{90.60} \\
  & Hard & \secondbest{32.00} & \secondbest{63.90} & \best{36.68} & \best{74.50} & 23.82 & 55.96 \\
\midrule
\multirow{2}{*}{Llama-3.1-8B}
  & Easy & \best{97.92} & 85.26 & 95.66 & \best{92.94} & \secondbest{96.68} & \secondbest{86.80} \\
  & Hard & \secondbest{23.56} & \secondbest{65.02} & \best{34.72} & \best{81.56} & 22.86 & 51.74 \\
\bottomrule
\end{tabular}
}
\caption{Pairwise accuracy (\%) on easy vs.\ hard pairs on \domain. \textbf{Bold}: best; \underline{underline}: second best, per LLM per difficulty level.}
\label{tab:pair_level_acc}
\end{table}

\subsection{Results on \domain}
\label{sec:domain_results}

% We observe that LLMs exhibit a foundational, albeit inconsistent, authority perception capability. 

% \subsubsection{Overall Performance} 
% Table~\ref{tab:domain_auth_fine} presents the main results on \domain under the fine-grained \liste setting. We observe that LLMs exhibit a foundational, albeit inconsistent, authority perception capability. 
% Larger models like Qwen3-32B and Gemini-3-Flash demonstrate reasonably strong perception ($\rho$ up to 0.75 and 0.80, respectively), indicating a developing understanding of source authority. However, smaller models and certain methods yield correlations as low as 0.20, suggesting this capability is not yet robustly developed across the board. 
% % Figure~\ref{fig:domain_auth_results} provides a visual summary for the open-source models.

% \begin{figure*}[t]
% \centering
% \includegraphics[width=\textwidth]{Sections/figure/fig_domain_auth_results.pdf}
% \caption{Spearman's $\rho$ correlation on \domain (fine-grained, 10-level) for open-source models. The left panel shows performance without text, and the right panel shows performance with text. Score-based methods generally outperform direct-output methods, but performance for relative judges degrades when text is introduced.}
% \label{fig:domain_auth_results}
% \end{figure*} 

Table~\ref{tab:domain_auth_fine}, Table \ref{tab:domain_auth_coarse}, and Table \ref{tab:pair_level_acc} present the main results on \domain under the fine-grained \liste setting, coarse-grained \liste setting, and \paire setting, respectively. 
% \heading{Comparison of Different Judges}
We can observe that:
(1) For the \liste setting, ListJudge and PairJudge with PointScore have better performance than other methods among all LLMs. PairJudge with PointScore achieves the best performance in most settings.  Compared to PointJudge, ListJudge, and PairJudge have more interaction among other domains before giving the absolute authority score. 
PointScore has better performance than ListRank on all settings and all LLMs. The reason may be that ListRank compels the model to make fine-grained, relative distinctions, producing sharper orderings—especially in near-tie cases. By contrast, PointScore assigns independent scalar scores and then sorts them, which lowers variance but smooths away subtle differences and weakens cross-item calibration. In short, ListRank trades difficulty for discriminative power, while PointScore favors stability at the cost of missed fine-grained distinctions.

% ListRank has better performance than PointScore on all settings and all LLMs. The reason may be that ListRank forces the LLMs to make fine-grained, relative distinctions among domains, especially in near-tie cases. This requires precise cross-item calibration and is sensitive to small stylistic or positional biases, leading to more order inversions and non-transitive preferences.  PointScore only asks the model to assign a scalar score to the current domain in the context of other domains. The final ranking is derived by sorting these scores, which reduces variance, smooths over small errors, and yields more consistent global calibration. 
% For Qwen3-32B (w/o text), PairJudge achieves the best overall performance ($\rho=0.7528$).  
% This pattern suggests that by requiring the model to first assign an intermediate numerical score on a fixed scale---rather than directly producing a permutation or preference decision---we effectively replicate the anchoring mechanism that makes Pointwise-Judge robust.
% Decoupling the comparative judgment from the final output generation allows the model to leverage its latent authority knowledge more reliably, bypassing the cognitive load imposed by direct ranking under simultaneous multi-document comparison. 
(2) Moreover, comparing the different LLMs on correlation performance increases monotonically with model size (8B $<$ 14B $<$ 32B) under all judges and evaluation settings, indicating that authority perception, as a nuanced dimension of world knowledge, benefits from increased model capacity. We also observe that Qwen3-8B frequently fails to follow output format constraints---returning free-form text instead of the required preference decision---causing parsing failures that substantially hurt its performance. Particularly under the PairJudge-PairRank setting, it performs considerably worse than same-sized models (Tables~\ref{tab:domain_auth_fine}~\&~\ref{tab:domain_auth_coarse}), which we attribute to its lack of instruction tuning.

(3) For the impact of textual information, adding text generally improves or maintains performance on all settings for PointJudge, while bringing a performance drop on most settings for ListJudge and PairJudge, especially under the \liste setting. The reasoning may be that authority is not equivalent to textual style, fluency, or narrative richness. In PairJudge/ListJudge, concatenating multiple texts multiplies length and heterogeneity, diluting attention and shifting the model toward surface features rather than authority. 
However, on \textit{hard-pairs} (small authority gap) under the \paire setting, adding text \textit{substantially improves} accuracy. This indicates that textual content can serve as a valuable compensatory signal precisely when structural authority cues are ambiguous. Moreover, the PointJudge has the worst performance. The reasoning may be that PointJudge tends to be concentrated, as shown in Appendix \ref{app:analysis_Juge}.

\begin{table*}[t]
\centering
\renewcommand{\arraystretch}{0.8}
\resizebox{\textwidth}{!}{%
\begin{tabular}{llcccccccccc}
\toprule
\multicolumn{2}{c}{} &
  \multicolumn{2}{c}{\textbf{PointJudge}} &
  \multicolumn{4}{c}{\textbf{ListJudge}} &
  \multicolumn{4}{c}{\textbf{PairJudge}} \\
\cmidrule(lr){3-4}\cmidrule(lr){5-8}\cmidrule(lr){9-12}
\multicolumn{2}{c}{} & & &
  \multicolumn{2}{c}{\textbf{ListRank}} &
  \multicolumn{2}{c}{\textbf{PointScore}} &
  \multicolumn{2}{c}{\textbf{PairRank}} &
  \multicolumn{2}{c}{\textbf{PointScore}} \\
\cmidrule(lr){5-6}\cmidrule(lr){7-8}\cmidrule(lr){9-10}\cmidrule(lr){11-12}
\textbf{Domain} & \textbf{Model} &
  $\rho$ & $\tau$ & $\rho$ & $\tau$ & $\rho$ & $\tau$ & $\rho$ & $\tau$ & $\rho$ & $\tau$ \\
\midrule
\multirow{5}{*}{\rotatebox[origin=c]{90}{\textbf{Basketball}}}
  & Qwen3-8B         & 74.93 & 61.26 & 65.38 & 54.93 & 72.53 & 59.32 & \secondbest{81.29} & \secondbest{68.07} & \best{86.70} & \best{73.90} \\
  & Qwen3-14B             & 80.26 & 66.10 & 69.37 & 60.94 & 76.42 & 65.29 & \secondbest{85.09} & \secondbest{71.97} & \best{85.89} & \best{73.07} \\
  & Qwen3-32B             & \secondbest{85.17} & \secondbest{70.98} & 64.22 & 57.53 & 75.33 & 64.96 & 82.25 & 69.22 & \best{85.94} & \best{73.66} \\
  & Llama-3-8B-Instruct   & 76.00 & 67.33 & 59.91 & 49.13 & 75.54 & 65.94 & \secondbest{83.54} & \secondbest{70.66} & \best{88.90} & \best{77.07} \\
  & Llama-3.1-8B-Instruct & \secondbest{82.55} & \secondbest{70.94} & 59.38 & 50.22 & 76.00 & 64.90 & 79.90 & 66.10 & \best{87.43} & \best{76.45} \\
\midrule
\multirow{5}{*}{\rotatebox[origin=c]{90}{\textbf{Movies}}}
  & Qwen3-8B         & 81.25 & 66.02 & 78.30 & 63.91 & \secondbest{82.40} & \secondbest{67.22} & 61.33 & 47.18 & \best{87.87} & \best{75.20} \\
  & Qwen3-14B             & 78.62 & 62.77 & \secondbest{87.70} & \secondbest{74.62} & 85.12 & 70.74 & 86.19 & 72.75 & \best{88.60} & \best{75.94} \\
  & Qwen3-32B             & 85.30 & 70.69 & \secondbest{87.76} & \secondbest{74.77} & 87.67 & 74.60 & 83.27 & 69.16 & \best{89.56} & \best{77.40} \\
  & Llama-3-8B-Instruct   & 78.24 & 62.77 & 75.06 & 60.60 & \best{84.91} & \best{71.50} & 79.04 & 64.74 & \secondbest{83.24} & \secondbest{69.66} \\
  & Llama-3.1-8B-Instruct & 81.33 & 66.37 & 80.69 & 66.28 & \best{84.38} & \best{71.11} & 70.00 & 55.42 & \secondbest{82.72} & \secondbest{69.32} \\
\midrule
\multirow{5}{*}{\rotatebox[origin=c]{90}{\textbf{Songs}}}
  & Qwen3-8B         & \best{46.56} & \best{34.96} & 34.78 & 25.87 & 42.71 & 31.84 & 10.57 & 7.78 & \secondbest{44.58} & \secondbest{33.52} \\
  & Qwen3-14B             & 47.19 & 35.31 & \best{51.75} & \best{38.64} & 49.02 & 36.64 & 47.62 & 35.52 & \secondbest{50.54} & \secondbest{37.75} \\
  & Qwen3-32B             & 49.72 & 37.10 & \secondbest{51.83} & 38.64 & \best{53.24} & \best{39.84} & 39.92 & 29.68 & 51.66 & \secondbest{38.78} \\
  & Llama-3-8B-Instruct   & \secondbest{49.92} & \secondbest{37.46} & 45.98 & 34.28 & \best{52.38} & \best{39.11} & 25.62 & 18.90 & 48.33 & 35.95 \\
  & Llama-3.1-8B-Instruct & 42.63 & 32.03 & \best{49.69} & \best{37.16} & \secondbest{49.59} & \secondbest{37.15} & 40.27 & 29.86 & 46.76 & 34.84 \\
\bottomrule
\end{tabular}
}
\caption{Results on \entity (fine-grained, 10-level, w/o text) across three domains reported as Spearman's $\rho$ (\%) and Kendall's $\tau$ (\%). \textbf{Bold}: best; \underline{underline}: second best, per LLM per domain.}
\label{tab:entity_auth_main}
\end{table*}

\subsection{Results on \entity}
% \subsubsection{Overall Performance}
Table~\ref{tab:entity_auth_main} and Table \ref{tab:entity_auth_coarse} (in the Appendix~\ref{app:coarse_grained_results}) show the performance of different methods on \entity benchmark and using fine- and coarse-labels, respectively.  
From the tables, we can find that: 
(1) The ListJudge and PairJudge methods, when combined with PointScore, perform better than other methods and exhibit similar performance trends. Moreover, the cross-LLMs patterns on \entity largely mirror those observed on \domain. 
(2) Performance varies considerably across the three domains. LLMs achieve notably stronger and more consistent results on Basketball and Movies compared to Songs. 
The Basketball and Movies in Wikipedia links have roughly five times as many links as Songs. 
So, the performance difference may stem from differences in the volume of domains in the pre-training data of LLMs. 
(3) Compared to the performance on \domain benchmark in Table~\ref{tab:domain_auth_fine}\& \ref{tab:domain_auth_coarse}, LLMs generally exhibit stronger and more consistent authority perception on \entity compared to \domain, particularly in the Basketball and Movies domains. 
For instance, Qwen3-32B achieves a Spearman's $\rho$ of 0.8594 on Basketball, higher than its best score of 0.7528 on \domain. This performance difference can be attributed to the nature of the authority signal: entity authority is often more concrete, objective, and well-documented in structured formats (e.g., player statistics, filmographies) within the training data, making it an easier concept for models to grasp compared to the more abstract notion of website authority.

\begin{figure}[t]
\centering
\includegraphics[width=\columnwidth]{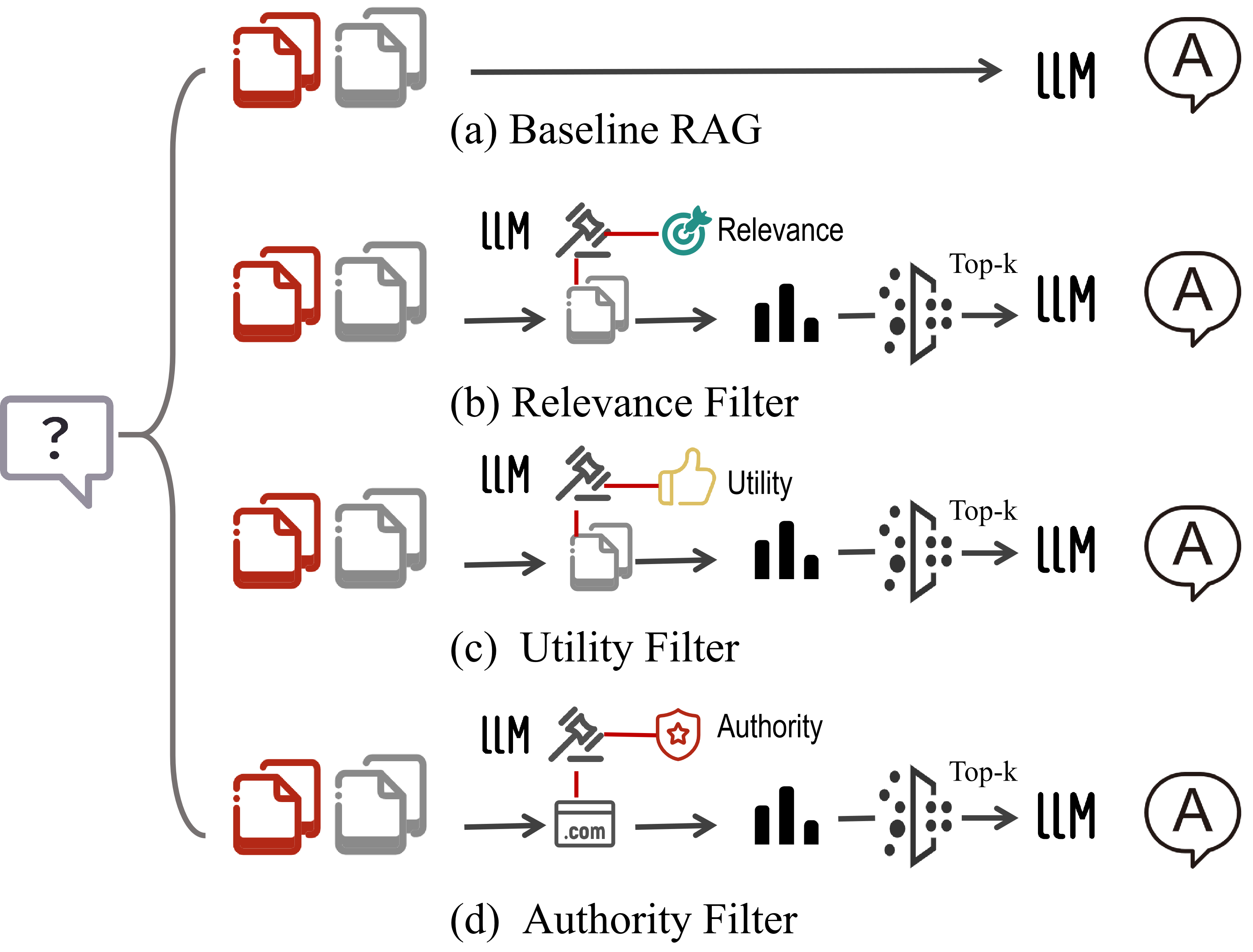}
% \caption{Illustration of the authority-aware RAG pipeline. All filtering methods share the same Listwise-PointScore paradigm; they differ only in the scoring criterion: (a) w/o filtering (Baseline RAG), (b) authority prompt without filtering, (c) LLM-based authority filtering, and (d) authority filtering combined with an explicit authority prompt at generation.}
\caption{
% Illustration of the authority-aware RAG pipeline. Documents are first retrieved and scored by the LLM judge; top-$k$ documents are then passed to the generator. The four strategies differ only in the scoring criterion used for filtering: (a)~\textbf{w/o Filter}: no filtering; (b)~\textbf{Relevance Filter}: documents are scored by relevance to the question; (c)~\textbf{Utility Filter}: documents are scored by their utility in generating a correct answer; (d)~\textbf{Authority Filter}: documents are scored solely by the authority of their source URL, without access to document content.
Authority-aware RAG pipeline: retrieved documents are LLM-scored; top-$k$ go to the generator. Filtering signals: (a)~\textbf{w/o Filter}; (b)~\textbf{Relevance Filter}; (c)~\textbf{Utility Filter}; (d)~\textbf{Authority Filter} (source URL only; no document content). All prompts in this section are provided in Appendix~\ref{sec:prompts}.
}
\label{fig:rag_strategies}
\end{figure}

\subsection{Authority-Aware RAG}
\label{sec:ragauth}
To evaluate the impact of authority-awareness on RAG, we compare several document filtering strategies on our proposed \rag. All strategies follow a unified pipeline: first, LLMs re-rank $N$ documents based on different criteria; then, the top-$k$ documents are passed to the final answer generation.  While our findings indicate that PairJudge-PointScore generally achieves the best performance in discerning authority, its high computational cost makes it impractical for real-time RAG applications. ListJudge, which scores all documents in a single pass, offers a more cost-effective trade-off.
For a fair comparison across criteria, all strategies use ListJudge inputs and the PointScore protocol to rank all documents, as shown in Figure~\ref{fig:rag_strategies}: 
\begin{itemize}[leftmargin=*,itemsep=2pt,topsep=3pt,parsep=0pt]
    \item \textbf{Baseline RAG:} No filtering is applied. 
    \item \textbf{Relevance Filter:} The document is scored based on its relevance to the query.
    \item \textbf{Utility Filter:} Following \citet{zhang2024large}, the document is scored based on its utility for answering the question. The model is first prompted to generate a pseudo-answer from the document list, and then scores each document based on its utility in answering the question. 
    \item \textbf{Authority Filter:}  Score each document based on the perceived authority of its source using the ListJudge-PointScore method, without considering the query content. 
\end{itemize}

% The prompts for the four strategies and the final prompts used to generate answers are provided in Appendix~\ref{sec:prompts}

% Table: RAGAuth
\begin{table}[t]
\centering
\renewcommand{\arraystretch}{0.95}
\small
\setlength{\tabcolsep}{8pt}
\begin{tabular}{lrccc}
\toprule
\textbf{Model} & $k$ &
  \textbf{Relevance} & \textbf{Utility} & \textbf{Authority} \\
\midrule
\multirow{4}{*}{\shortstack[l]{Qwen3\\-8B}}
  & -- & \multicolumn{3}{c}{w/o Filter: 53.33} \\
  & 1  & 48.33 & 51.67 & \best{66.67} \\
  & 3  & 51.67 & 58.33 & \best{69.16} \\
  & 5  & 51.67 & 58.33 & \best{69.16} \\
\midrule
\multirow{4}{*}{\shortstack[l]{Qwen3\\-14B}}
  & -- & \multicolumn{3}{c}{w/o Filter: 58.33} \\
  & 1  & 51.67 & 60.00 & \best{76.67} \\
  & 3  & 45.00 & 66.67 & \best{75.00} \\
  & 5  & 45.00 & 66.67 & \best{73.33} \\
\midrule
\multirow{4}{*}{\shortstack[l]{Qwen3\\-32B}}
  & -- & \multicolumn{3}{c}{w/o Filter: 55.00} \\
  & 1  & 63.33 & 65.00 & \best{70.00} \\
  & 3  & 58.33 & 63.33 & \best{68.33} \\
  & 5  & 60.00 & 60.00 & \best{70.00} \\
\midrule
\multirow{4}{*}{\shortstack[l]{Llama\\-3-8B}}
  & -- & \multicolumn{3}{c}{w/o Filter: 50.83} \\
  & 1  & 37.50 & 49.17 & \best{64.17} \\
  & 3  & 41.67 & 52.50 & \best{64.17} \\
  & 5  & 41.67 & 52.50 & \best{64.17} \\
\midrule
\multirow{4}{*}{\shortstack[l]{Llama\\-3.1-8B}}
  & -- & \multicolumn{3}{c}{w/o Filter: 57.50} \\
  & 1  & 42.50 & 51.67 & \best{63.34} \\
  & 3  & 55.00 & 48.33 & \best{71.76} \\
  & 5  & 55.00 & 48.33 & \best{71.76} \\
\bottomrule
\end{tabular}
\caption{Answer accuracy (\%) on \rag across four models under different filtering strategies and top-$k$ settings ($k \in \{1,3, 5\}$). \textbf{Bold}: best per LLM per $k$.}
\label{tab:rag_results}
\end{table}

% \begin{figure}[t]
% \centering
% \includegraphics[width=\columnwidth]{Sections/figure/Authority_aware_RAG.png}
% \caption{Illustration of the authority-aware RAG pipeline. All filtering methods share the same listwise-input, pointwise-scoring framework; they differ only in the scoring criterion: (a) w/o filtering (Baseline RAG), (b) authority prompt without filtering, (c) LLM-based authority filtering, and (d) authority filtering combined with an explicit authority prompt at generation.}
% \label{fig:rag_strategies}
% \end{figure}

\heading{RAG Performance}
As shown in Table~\ref{tab:rag_results}, we can find that: 
(1) All models demonstrate performance improvement when incorporating filtering methods based on different criteria. 
(2) Incorporating authority signals during filtering has the best RAG performance improvement across all strategies. This underscores the central role of source authority in enhancing answer accuracy and reducing susceptibility to low-credibility evidence. 
\section{Conclusion}
\label{sec:conclusion}
In this work, we introduced \benchname, a comprehensive benchmark designed to evaluate the authority perception capabilities of LLMs. 
Through extensive experiments on our three datasets, \domain, \entity, and \rag, our key findings are as follows: (1) ListJudge and PairJudge have the strongest performance compared to other methods. Moreover, the optimal method for eliciting authority judgments is context-dependent. (2) Finally, we demonstrated the practical value of authority perception in a downstream RAG task, where authority-aware filtering achieved the best answer accuracy compared to other methods. 
Our work provides a comprehensive benchmark for understanding and harnessing the authority perception of LLMs, a crucial step towards building more reliable and trustworthy AI systems.

% \clearpage
\section{Limitations}
\label{sec:limitations}
Our study has several limitations that offer avenues for future research. 
First is our reliance on PageRank as the primary ground truth for domain authority. While PageRank is a well-established, objective, and scalable proxy for authority based on the web's link structure, it is not without its flaws. It can be slow to adapt to the rise of new authoritative sources and may not fully capture the nuanced, topic-specific authority that a human expert might perceive. For instance, a niche but highly respected academic blog might have a lower PageRank than a mainstream news aggregator, yet be more authoritative on its specific topic. Moreover, certain highly authoritative official websites may receive disproportionately low PageRank scores due to sparse inbound links. We analyze this issue in detail in Appendix~\ref{app:analysis_Juge}. Future work can address this gap by integrating more dynamic, context-aware, or topic-specific authority metrics into the evaluation framework.
Second, the scale of our \rag dataset is modest due to the manual cost. While the 120 questions are sufficient to demonstrate the potential of authority-aware filtering, a larger and more diverse set of questions and documents would be needed to draw more robust conclusions about its general applicability and to analyze the performance on different types of questions. 

% Finally, our analysis of textual impact primarily focused on the presence or absence of full webpage text. A more fine-grained analysis, such as examining the impact of specific textual features (e.g., author information, publication date, reference lists) or using summary-based content, could yield deeper insights into how LLMs weigh different types of evidence when judging authority.

% Fourth, the scale of our \rag dataset is modest. While the 125 questions are sufficient to demonstrate the potential of authority-aware filtering, a larger and more diverse set of questions and documents would be needed to draw more robust conclusions about its general applicability and to analyze the performance on different types of questions.

\section{Ethics Statement} 
In our work, the data and models we use are publicly available. We have transformed the original datasets to construct our \benchname. Both the queries and documents for the \rag are obtained from public web pages collected by authors and do not contain personal or sensitive information. 
To advance the community, we will release the benchmark we constructed along with the experimental code.

% \clearpage
\bibliography{custom}
\appendix
\clearpage
\newpage
\section{Prompts}
\label{sec:prompts}

\subsection*{Prompts for DomainAuth}

% -----------------------------------------------------------
% \subsubsection*{Pointwise --- w/o text}

\begin{promptbox}{A. DomainAuth / PointJudge (w/o text)}
\begin{pblock}
You are a web analysis expert. Please score the source authority of the given domain based on the following criteria, with a rating range of 0 to 9.

Source authority refers to the credibility, professionalism, official certification, or public recognition of the domain itself.

\textbf{DOMAIN}: [Domain name]

Based on the above criteria, rate the source authority of this DOMAIN (0--9 points).

Output only: \{"score": X\} where X is your evaluated score (0, 1, 2, 3, 4, 5, 6, 7, 8, or 9). Do not add any explanation.

Your answer:
\end{pblock}
\end{promptbox}

\vspace{0.8em}

% -----------------------------------------------------------
% \subsubsection*{Pointwise --- w/ text}

\begin{promptbox}{B. DomainAuth / PointJudge (w/ text)}
\begin{pblock}
You are a web analysis expert. Please score the source authority of the given domain based on the following criteria, with a rating range of 0 to 9.

Source authority refers to the credibility, professionalism, official certification, or public recognition of the domain itself.

\textbf{DOMAIN}: [Domain name]

\textbf{Webpage Text}: [Webpage Text]

Based on the domain name, their web content and the above criteria, rate the source authority of this DOMAIN (0--9 points).

Output only: \{"score": X\} where X is your evaluated score (0, 1, 2, 3, 4, 5, 6, 7, 8, or 9). Do not add any explanation.

Your answer:
\end{pblock}
\end{promptbox}

\vspace{0.8em}

% -----------------------------------------------------------
% \subsubsection*{Pairwise --- Direct Preference (w/o text)}

\begin{promptbox}{C. DomainAuth / PairJudge-PairRank (w/o text)}
\begin{pblock}
You are a web analysis expert. Please determine which of the two given domains is more authoritative as an information source based on the following criteria.

Source authority refers to the credibility, professionalism, official certification, or public recognition of the domain itself.

Consider both domains together when making the comparison.

\textbf{DOMAINS}:

Domain A: [Domain\_A]

Domain B: [Domain\_B]

Which domain is more authoritative? Only return `A' or `B'. Do not add any explanation.

Your answer:
\end{pblock}
\end{promptbox}

\vspace{0.8em}

% -----------------------------------------------------------
% \subsubsection*{Pairwise --- Direct Preference (w/ text)}

\begin{promptbox}{D. DomainAuth / PairJudge-PairRank (w/ text)}
\begin{pblock}
You are a web analysis expert. Please determine which of the two given domains is more authoritative as an information source based on the following criteria.

Source authority refers to the credibility, professionalism, official certification, or public recognition of the domain itself and its content. 

Consider both domains and their contents together when making the comparison.

\textbf{DOMAINS \& CONTENTS}:\\
Domain A: [Domain A]\\
Content A: [Text A]\\
Domain B: [Domain B]\\
Content B: [Text B]

Which domain is more authoritative? Only return `A' or `B'. Do not add any explanation.

Your answer:
\end{pblock}
\end{promptbox}

\vspace{0.8em}

% -----------------------------------------------------------
% \subsubsection*{Pairwise --- Score-Based}

\begin{promptbox}{E. DomainAuth / PairJudge-PointScore}
\begin{pblock}
You are a web analysis expert. Please score the source authority of each of the two given domains based on the following criteria, with a rating range of 0 to 9.

Source authority refers to the credibility, professionalism, official certification, or public recognition of the domain itself.

Consider both domains together when assigning scores to ensure consistency.

\textbf{DOMAINS}:\\
Domain A: [Domain A]\\
Domain B: [Domain B]

Your output MUST be a single, valid JSON object in the format: \{"A": X, "B": X\}, where each X is an integer score from 0 to 9.

Example format: \{"A": 7, "B": 4\}. Do not add any explanation.

Your answer:
\end{pblock}
\end{promptbox}

\vspace{0.8em}

% -----------------------------------------------------------
% \subsubsection*{Listwise --- Direct Permutation (w/ text)}
\newpage
\begin{promptbox}{F. DomainAuth / ListJudge-ListRank (w/ text)}
\begin{pblock}
You are a web analysis expert that can rank domains based on their authority.

I will provide you with 10 domains and their web contents, each indicated by a numerical identifier []. Please rank the domains in the provided list based on the following criteria.

Source authority refers to the credibility, professionalism, official certification, or public recognition of the domain itself.

\textbf{DOMAINS}:\\
{[0]} [Domains\_0]: [Text\_0] \\
{[1]} [Domains\_1]: [Text\_1] \\
...

Rank the 10 domains above based on their authority (highest to lowest).
All the domains should be included and listed using identifiers, in descending order of authority.

The output format should be [] > [], e.g., [4] > [0].

Only respond with the ranking results. Do not add any explanation.

Your answer:
\end{pblock}
\end{promptbox}

\vspace{0.8em}

% -----------------------------------------------------------
% \subsubsection*{Listwise --- Score-Based (w/o text)}

\begin{promptbox}{G. DomainAuth / ListJudge-PointScore(w/o text)}
\begin{pblock}
You are a web analysis expert. Please score the source authority of each domain in the provided list based on the following criteria, with a rating range of 0 to 9.

Source authority refers to the credibility, professionalism, official certification, or public recognition of the domain itself.

Consider all domains together when assigning scores to ensure consistency and relative calibration.

\textbf{DOMAINS}:\\
{[0]} [Domains\_0]: [Text\_0]\\
{[1]} [Domains\_1]: [Text\_1]\\
...

Your output MUST be a single, valid JSON object. The JSON object must map each domain's numerical identifier (as a string) to its integer score (0--9).

Example format: \{"0": 7, "1": 4, ..., "9": 8\}. Do not add any explanation.

Your answer:
\end{pblock}
\end{promptbox}

% ============================================================
%  SECTION 2: EntityAuth
% ============================================================
\newpage
\section*{Prompts for EntityAuth}

% -----------------------------------------------------------
% \subsubsection*{Pointwise}

\begin{promptbox}{H. EntityAuth / PointJudge}
\begin{pblock}
You are a Wikipedia notability expert and a music industry expert. Please score the authority of the given song entity based on the following criteria, with a rating range of 0 to 9.

Entity authority refers to the public recognition, fame, historical significance, or widespread knowledge of the entity (person, place, organization, etc.).

\textbf{QUESTION CONTEXT}: [Question]

\textbf{TARGET ENTITY}: [Question\_entity]

Based on the question context and the above criteria, rate the authority of this ENTITY (0--9 points).

Output only: \{"score": X\} where X is your evaluated score (0, 1, 2, 3, 4, 5, 6, 7, 8, or 9). Do not add any explanation.

Your answer:
\end{pblock}
\end{promptbox}

\vspace{0.8em}

% -----------------------------------------------------------
% \subsubsection*{Pairwise --- Direct Preference}

\begin{promptbox}{I. EntityAuth / PairJudge-PairRank}
\begin{pblock}
You are a Wikipedia notability expert and a music industry expert. Please determine which of the two given entities is more authoritative based on the following criteria.

Entity authority refers to the public recognition, fame, historical significance, or widespread knowledge of the entity (person, place, organization, etc.).

Consider both entities together when assigning scores to ensure consistency.

\textbf{QUESTION CONTEXT}: These entities are the subjects of questions like `Who is the performer of the song ...'

\textbf{TARGET ENTITIES}:\\
Entity A: ``[Entity\_A]''\\
Entity B: ``[Entity\_B]''

Which entity is more authoritative? Only return `A' or `B'. Do not add any explanation.

Your answer:
\end{pblock}
\end{promptbox}

\vspace{0.8em}

% -----------------------------------------------------------
% \subsubsection*{Pairwise --- Score-Based}
\newpage
\begin{promptbox}{J. EntityAuth / PairJudge-PointScore}
\begin{pblock}
You are a Wikipedia notability expert and a music industry expert. Please score the entity authority of each of the two given song entities based on the following criteria, with a rating range of 0 to 9.

Entity authority refers to the public recognition, fame, historical significance, or widespread knowledge of the entity (person, place, organization, etc.).

Consider both entities together when assigning scores to ensure consistency.

\textbf{QUESTION CONTEXT}: These entities are the subjects of questions like `Who is the performer of the song...'

\textbf{TARGET ENTITIES}:\\
Entity A: ``[Entity\_A]''\\
Entity B: ``[Entity\_B]''

Your output MUST be a single, valid JSON object in the format: \{"A": X, "B": Y\}, where X and Y are integer scores from 0 to 9.

Example format: \{"A": 7, "B": 4\}. Do not add any explanation.

Your answer:
\end{pblock}
\end{promptbox}

\vspace{0.8em}

% -----------------------------------------------------------
% \subsubsection*{Listwise --- Direct Permutation}

\begin{promptbox}{K. EntityAuth / ListJudge-ListRank}
\begin{pblock}
You are a Wikipedia notability expert and a music industry expert.
I will provide you with 10 song entities, each indicated by a numerical identifier [].
Please rank the song entities in the provided list based on the following criteria.

Entity authority refers to the public recognition, fame, historical significance, or widespread knowledge of the entity (person, place, organization, etc.).

Consider all entities together when assigning scores to ensure consistency and relative calibration.

\textbf{QUESTION CONTEXT}: These entities are the subjects of questions like `Who is the performer of the song...'

\textbf{TARGET ENTITIES}:\\
{[0]} [Question\_entity\_0]\\
{[1]} [Question\_entity\_1]\\
...

Rank the 10 song entities above based on their authority (highest to lowest).
All the song entities should be included and listed using identifiers, in descending order of authority.

The output format should be [] > [], e.g., [4] > [0].

Only respond with the ranking results, do not add any explanation.

Your answer:
\end{pblock}
\end{promptbox}

\vspace{0.8em}

% -----------------------------------------------------------
% \subsubsection*{Listwise --- Score-Based}
\newpage
\begin{promptbox}{L. EntityAuth / ListJudge-PointScore}
\begin{pblock}
You are a Wikipedia notability expert and a music industry expert. Please score the entity authority of each song entity in the provided list based on the following criteria, with a rating range of 0 to 9.

Entity authority refers to the public recognition, fame, historical significance, or widespread knowledge of the entity (person, place, organization, etc.).

Consider all entities together when assigning scores to ensure consistency and relative calibration.

\textbf{QUESTION CONTEXT}: These entities are the subjects of questions like `Who is the performer of the song...'

\textbf{TARGET ENTITIES}:\\
{[0]} [Question\_entity\_0] \\
{[1]} [Question\_entity\_1] \\
...

Your output MUST be a single, valid JSON object. The JSON object must map each entity's numerical identifier (as a string) to its integer score (0--9).

Example format: \{"0": 7, "1": 4, ..., "9": 8\}. Do not add any explanation.

Your answer:
\end{pblock}
\end{promptbox}

% ============================================================
%  SECTION 3: RAGAuth
% ============================================================
% \newpage
\subsection*{Prompts for RAGAuth}

% -----------------------------------------------------------
% \subsubsection*{Condition a: No Authority Information}

\begin{promptbox}{Strategy a: w/o Filter / Generate Final Answer}
\begin{pblock}
You are given a question and several pieces of documents. Your task is to analyze the documents and provide a concise answer to the question based solely on the provided reference documents.

\textbf{Question}: [Question]

\textbf{Reference Documents}:\\
Document [0]\\
Source URL: [URL]\\
Content: [Doc\_text]\\
...

Please answer with ONLY `yes' or `no' based on the documents above. Do not add any explanation.

Your answer:
\end{pblock}
\end{promptbox}
\vspace{0.8em}

% -----------------------------------------------------------
% \subsection*{Condition 2: With Source Authority Guidance}
\newpage
\begin{promptbox}{Strategy b: Relevance Filter --- Scoring Prompt}
\begin{pblock}
You are a web analysis expert. Please score the relevance of each document to the question with a rating range of 0 to 9, where 0 is the lowest and 9 is the highest.\\ 
Consider all documents together when assigning scores to ensure consistency and relative calibration.

\textbf{QUESTION}: [Question]

\textbf{DOCUMENTS}:\\\
[0]  [URL]: [Doc\_text]\\\ 
[1]  [URL]: [Doc\_text]\\
...

Your output MUST be a single, valid JSON object. The JSON object must map each document's numerical identifier (as a string) to its integer score (0--9).\\ 
Example format: \{"0": 7, "1": 4, ..., "9": 8\}. Do not add any explanation.\\ 

Your answer:
\end{pblock}
\end{promptbox}
\vspace{0.8em}

\begin{promptbox}{Strategy c: Utility Filter --- Scoring Prompt}
\begin{pblock}
You are a web analysis expert. Please score the utility of each document in answering the question based on the following criteria, with a rating range of 0 to 9, where 0 is the lowest and 9 is the highest. \\ 
The passage has utility in answering the question, meaning that the passage not only is relevant to the question, but also is useful in generating a correct, reasonable, and perfect answer to the question.\\ 
Consider all documents together when assigning scores to ensure consistency and relative calibration.

\textbf{QUESTION}: [Question]

\textbf{DOCUMENTS}:\\\
[0]: [URL]: [Doc\_text]\\\
[1]: [URL]: [Doc\_text]\\
...

Step 1: Generate a pseudo-answer to the question based on all the documents.

Step 2: Based on the pseudo-answers, rate each document's utility in answering the question on a scale of 0 to 9.

Your output MUST be a single, valid JSON object with two keys:
\begin{itemize}[noitemsep,topsep=2pt]
  \item \texttt{"answers"}: maps each document's numerical identifier (as a string) to its pseudo-answer
  \item \texttt{"scores"}: maps each document's numerical identifier (as a string) to its integer score (0--9)
\end{itemize}
Example format: \{"answers": \{"0": "...", "1": "...", ...\}, "scores": \{"0": 7, "1": 4, ...\}\}. Do not add any explanation.\\
Your answer:
\end{pblock}
\end{promptbox}
\vspace{0.8em}

\begin{promptbox}{Strategy d: Authority Filter --- Scoring Prompt}
\begin{pblock}
You are a web analysis expert. Please score the source authority of each source URL in the provided list based on the following criteria, with a rating range of 0 to 9, where 0 is the lowest and 9 is the highest.
Source authority refers to the credibility, professionalism, official certification, or public recognition of the domain itself and its content.
Consider all URLs together when assigning scores to ensure consistency and relative calibration.

\textbf{SOURCES}:\\\
[0]: [URL]\\\
[1]: [URL]\\
...

Your output MUST be a single, valid JSON object. The JSON object must map each domain's numerical identifier (as a string) to its integer score (0--9).
Example format: \{"0": 7, "1": 4, ..., "9": 8\}. Do not add any explanation.\\
Your answer:
\end{pblock}
\end{promptbox}
\vspace{0.8em}

\begin{table*}[t]
\centering
\resizebox{\textwidth}{!}{%
\begin{tabular}{llcccccccccc}
\toprule
\multicolumn{2}{c}{} &
  \multicolumn{2}{c}{\textbf{PointJudge}} &
  \multicolumn{4}{c}{\textbf{ListJudge}} &
  \multicolumn{4}{c}{\textbf{PairJudge}} \\
\cmidrule(lr){3-4}\cmidrule(lr){5-8}\cmidrule(lr){9-12}
\multicolumn{2}{c}{} & & &
  \multicolumn{2}{c}{\textbf{ListRank}} &
  \multicolumn{2}{c}{\textbf{PointScore}} &
  \multicolumn{2}{c}{\textbf{PairRank}} &
  \multicolumn{2}{c}{\textbf{PointScore}} \\
\cmidrule(lr){5-6}\cmidrule(lr){7-8}\cmidrule(lr){9-10}\cmidrule(lr){11-12}
\textbf{Domain} & \textbf{Model} &
  $\rho$ & $\tau$ & $\rho$ & $\tau$ & $\rho$ & $\tau$ & $\rho$ & $\tau$ & $\rho$ & $\tau$ \\
\midrule
\multirow{5}{*}{\rotatebox[origin=c]{90}{\textbf{Basketball}}}
  & Qwen3-8B         & \secondbest{70.97} & \secondbest{60.04} & 50.60 & 44.77 & 66.64 & 56.60 & 29.52 & 25.73 & \best{84.19} & \best{74.97} \\
  & Qwen3-14B             & 78.08 & 67.70 & 70.10 & 64.93 & 71.52 & 63.60 & \best{86.50} & \best{78.39} & \secondbest{83.80} & \secondbest{75.04} \\
  & Qwen3-32B             & 79.67 & 68.93 & 62.53 & 60.06 & 73.39 & 65.96 & \best{84.76} & \best{76.81} & \secondbest{83.94} & \secondbest{76.22} \\
  & Llama-3-8B-Instruct   & 73.76 & 67.14 & 65.38 & 58.05 & 68.86 & 62.50 & \secondbest{83.91} & \secondbest{76.56} & \best{87.46} & \best{80.23} \\
  & Llama-3.1-8B-Instruct & 78.28 & 69.31 & 57.09 & 51.54 & 76.34 & 68.21 & \secondbest{78.53} & \secondbest{70.60} & \best{86.52} & \best{77.91} \\
\midrule
\multirow{5}{*}{\rotatebox[origin=c]{90}{\textbf{Movies}}}
  & Qwen3-8B         & 66.25 & 55.08 & 51.61 & 45.30 & \secondbest{71.94} & \secondbest{60.96} & 48.32 & 40.90 & \best{86.75} & \best{78.75} \\
  & Qwen3-14B             & 75.17 & 63.49 & 88.56 & 80.64 & 82.07 & 71.12 & \best{89.03} & \best{81.72} & \secondbest{88.77} & \secondbest{80.89} \\
  & Qwen3-32B             & 76.49 & 65.26 & \secondbest{88.42} & \secondbest{80.66} & 84.31 & 74.11 & 85.90 & 77.85 & \best{90.27} & \best{83.40} \\
  & Llama-3-8B-Instruct   & 76.49 & 65.26 & 75.77 & 65.11 & 76.22 & 65.80 & \secondbest{80.84} & \secondbest{71.60} & \best{87.42} & \best{79.68} \\
  & Llama-3.1-8B-Instruct & 58.45 & 48.15 & 78.30 & 68.03 & \secondbest{80.09} & \secondbest{69.60} & 67.35 & 57.21 & \best{80.84} & \best{71.07} \\
\midrule
\multirow{5}{*}{\rotatebox[origin=c]{90}{\textbf{Songs}}}
  & Qwen3-8B         & \best{42.51} & \best{34.34} & 23.80 & 19.45 & 30.11 & 24.32 & 24.58 & 19.93 & \secondbest{41.01} & \secondbest{33.40} \\
  & Qwen3-14B             & 45.67 & 37.09 & \best{51.76} & \best{42.34} & 47.28 & 38.39 & 48.89 & 39.92 & \secondbest{50.77} & \secondbest{41.40} \\
  & Qwen3-32B             & 47.71 & 38.80 & \best{52.25} & \best{42.79} & 50.50 & 41.07 & 39.60 & 32.23 & \secondbest{51.42} & \secondbest{42.16} \\
  & Llama-3-8B-Instruct   & \secondbest{48.41} & \secondbest{39.85} & 45.27 & 36.88 & 43.74 & 35.51 & 40.20 & 32.71 & \best{49.65} & \best{40.55} \\
  & Llama-3.1-8B-Instruct & 40.46 & 32.83 & 45.36 & 36.97 & \secondbest{46.09} & \secondbest{37.75} & 19.50 & 15.64 & \best{48.41} & \best{39.54} \\
\bottomrule
\end{tabular}
}
\caption{Results on \entity (coarse-grained, 5-level, w/o text) across three domains reported as Spearman's $\rho$ and Kendall's $\tau$ (\%). \textbf{Bold}: best; \underline{underline}: second best, per model per domain.} 
\label{tab:entity_auth_coarse}
\end{table*}

\section{Coarse-Grained Results}
\label{app:coarse_grained_results}
This section shows the results of \entity under coarse-grained, as shown in Table \ref{tab:entity_auth_coarse}. 

\section{An Example from RAGAuth Dataset}
\begin{figure}[t!]
\centering
\includegraphics[width=\columnwidth]{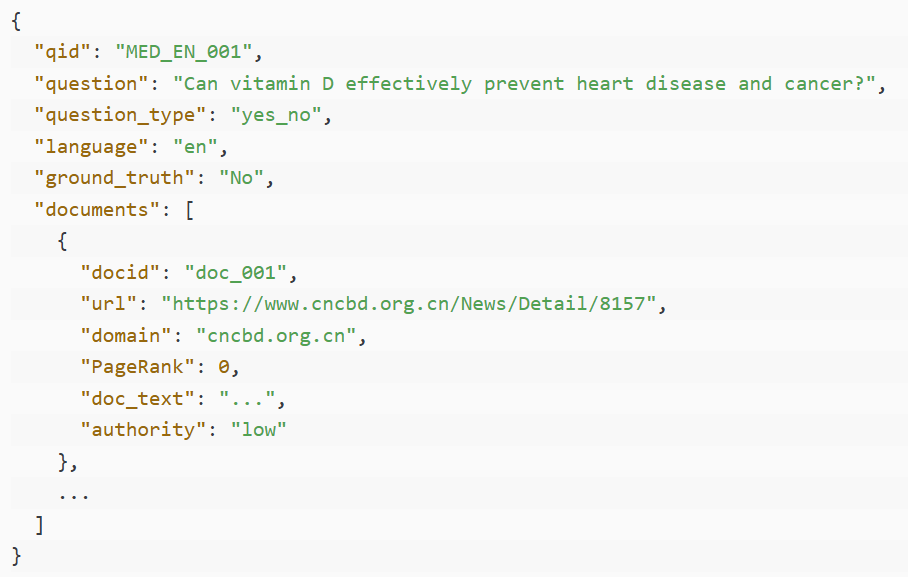}
\caption{An example from the \rag dataset. Each instance includes a yes/no question, the ground-truth answer, and a list of 10 retrieved documents with their source URL, domain, PageRank score, and text snippet. The task is to generate a correct answer based on this information.}
\label{fig:rag_data_example}
\end{figure}

Figure \ref{fig:rag_data_example} shows an example from RAGAuth dataset. 

% \begin{table*}[t]
% \centering
% \small
% \setlength{\tabcolsep}{4pt} % Reduce column spacing
% \label{tab:benchmark_overview}
% \begin{tabular}{llcccccc}
% \toprule
% \textbf{Sub-task} & \textbf{Domain} & \textbf{Authority Proxy} & \textbf{Score Scale} & \textbf{Corpus} & \textbf{Listwise (N=5/10)}$^\dagger$ & \textbf{Pairwise (Hard/Easy)}$^\ddagger$ \\
% \midrule
% \domain & Web Domains & PageRank & 0--9 / 0--4 & 10,000 & 4.5k / 4.5k & 5k / 5k \\
% \midrule
% \multirow{3}{*}{\entity} & Basketball & \multirow{3}{*}{Sitelinks} & \multirow{3}{*}{0--9} & 10,000 & -- / 4.5k & -- \\
% & Movies & & & 10,000 & -- / 4.5k & -- \\
% & Songs & & & 2,000 & -- / 4.5k & -- \\
% \midrule
% \rag & Yes/No QA & -- & N/A & 120 Queries & \multicolumn{2}{c}{4 conditions $\times$ 120 Queries} \\
% \bottomrule
% \end{tabular}
% \caption{Overview of the \benchname. \textbf{Corpus} is the number of unique items. For \domain, \textbf{Listwise} instances are split between $N{=}5$ and $N{=}10$ formats; \textbf{Pairwise} instances are split between hard ($0<\Delta\text{PageRank}{\leq}2$) and easy ($\Delta\text{PageRank}{\geq}5$) pairs. EntityAuth uses only $N{=}10$ lists and does not include a Pairwise setting. RAGAuth is evaluated under 4 experimental conditions.}
% \label{tab:data_statistics}
% \end{table*}

\begin{table*}[t]
\centering
\small
\setlength{\tabcolsep}{4pt} % Reduce column spacing
\label{tab:benchmark_overview}
\begin{tabular}{llcccccc}
\toprule
\textbf{Sub-task} & \textbf{Domain} & \textbf{Authority Proxy} & \textbf{Score Scale} & \textbf{Corpus} & \textbf{Listwise (N=5/10)}$^\dagger$ & \textbf{Pairwise (Hard/Easy)}$^\ddagger$ \\
\midrule
\domain & Web Domains & PageRank & 0--9 / 0--4 & 10,000 & 4.5k / 4.5k & 5k / 5k \\
\midrule
\multirow{3}{*}{\entity} & Basketball & \multirow{3}{*}{Sitelinks} & \multirow{3}{*}{0--9} & 10,000 & -- / 4.5k & -- \\
& Movies & & & 10,000 & -- / 4.5k & -- \\
& Songs & & & 2,000 & -- / 4.5k & -- \\
\midrule
\multirow{3}{*}{\rag} & Medical & \multirow{3}{*}{--} & \multirow{3}{*}{N/A} & \multirow{3}{*}{\makecell{120 Queries }} & \multicolumn{2}{c}{\multirow{3}{*}{10 docs/query}} \\
& Science & & & & \multicolumn{2}{c}{} \\
& Current Affairs & & & & \multicolumn{2}{c}{} \\
\bottomrule
\end{tabular}
\caption{Overview of the \benchname. \textbf{Corpus} is the number of unique items. $^\dagger$\textbf{Listwise} instances are split between $N{=}5$ and $N{=}10$ formats. $^\ddagger$\textbf{Pairwise} instances are split between hard ($0<\Delta\text{PageRank}{\leq}2$) and easy ($\Delta\text{PageRank}{\geq}5$) pairs. EntityAuth uses only $N{=}10$ lists and does not include a Pairwise setting. For RAGAuth, each query is paired with 10 retrieved webpage documents.}
\label{tab:data_statistics}
\end{table*}

\section{Benchmark Statistics}
Table \ref{tab:data_statistics} shows the statistics details of \benchname. 
% TABLE 1: BENCHMARK OVERVIEW

% \section{Coarse-Grained vs. Fine-Grained Comparison}
% In this section, we re-evaluate our main findings under a coarse-grained (5-level) setting. 

% As shown in Appendix~\ref{app:coarse_grained_results} (Table~\ref{tab:domain_auth_coarse}), the core patterns observed on \domain hold true. First, the Pointwise-Judge remains robust to the inclusion of text, while relative judges like Listwise-Direct still suffer a significant performance drop (e.g., for Qwen3-8B-Base, $\rho$ drops from 0.6560 to 0.2781). This confirms that the conflict between authority and content style signals is a fundamental issue for direct relative judgments, regardless of label granularity.

% Similarly, the analysis on \entity (coarse-grained results in Appendix~\ref{app:coarse_grained_results}, Table~\ref{tab:entity_auth_coarse}) shows that the optimal direct output method is still domain-dependent. For instance, in the Basketball domain, Pairwise-Direct continues to be the strongest method for the Qwen3 family (e.g., Qwen3-14B, $\rho=0.865$), while in the more subjective Movies and Songs domains, Listwise-Direct is preferred. The Llama3 family also maintains its preference for the Pointwise-Judge. 

\begin{figure}[t]
\centering
\includegraphics[width=\columnwidth]{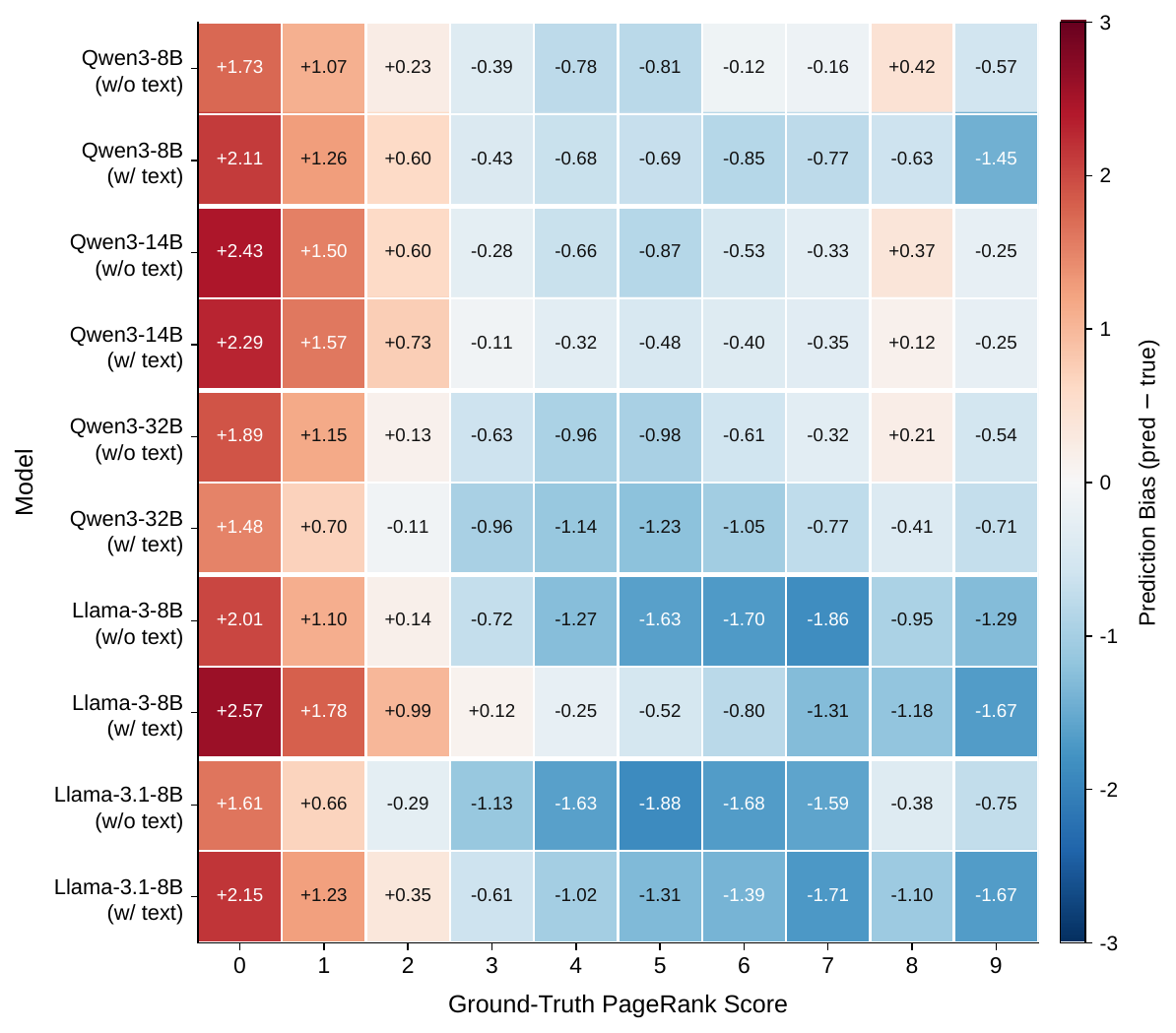}
\caption{Heatmap of prediction bias (predicted - true score) on \domain for the Pointwise-Judge. Red cells indicate overestimation, while blue cells indicate underestimation. Most models tend to overestimate low-authority sources and underestimate high-authority ones.}
\label{fig:scoring_bias_heatmap}
\end{figure}

\section{Analysis of Scoring Bias}
\label{app:analysis_Juge}
In this section, we analyze the overall scoring behavior of the PointJudge method and its discrepancies against the ground-truth authority proxy. Figure~\ref{fig:scoring_bias_heatmap} illustrates the bias in the average predicted score for each true authority score (0-9) on \domain. A consistent trend is observed across all models: they tend to overestimate the authority of low-authority sources while underestimating that of high-authority sources to a milder degree. Such a “regression to the mean” phenomenon reveals that, although PointJudge achieves competitive ranking performance, there still exist noticeable discrepancies between its predicted scores and the true authority proxy. 

To further investigate this bias, we perform a fine-grained analysis on three typical domain categories: government (\texttt{.gov}), education (\texttt{.edu}), and social media. Table~\ref{tab:domain_category_bias} reports the mean scoring bias for each category
% 是否需要展示我们分析了哪些主域名
% ------------------------------------------------------------------
% Table: Domain Category Scoring Bias (single-column, compact)
% ------------------------------------------------------------------
\begin{table}[t]
\centering
\small
\setlength{\tabcolsep}{2.8pt}
\caption{Mean predicted score and scoring bias across three domain
  categories. \textit{Bias} $=$ avg.\ (predicted $-$ true PageRank).
  True means: Gov\,5.68 ($N$\!=\!378); Edu\,5.64 ($N$\!=\!160);
  Social\,6.00 ($N$\!=\!34).}
\label{tab:domain_category_bias}
\begin{tabular}{@{}llcccccc@{}}
\toprule
& & \multicolumn{2}{c}{\textbf{.gov}} &
    \multicolumn{2}{c}{\textbf{.edu}} &
    \multicolumn{2}{c}{\textbf{Social}} \\
\cmidrule(lr){3-4}\cmidrule(lr){5-6}\cmidrule(lr){7-8}
\textbf{Model} & \textbf{Ctx.}
  & \small Pred & \small Bias
  & \small Pred & \small Bias
  & \small Pred & \small Bias \\
\midrule
\multirow{2}{*}{Qwen3-14B}
  & w/o & 8.97 & {+3.28} & 8.84 & {+3.20} & 8.44 & {+2.44} \\
  & w/  & 8.39 & {+2.70} & 8.25 & {+2.61} & 7.59 & {+1.59} \\
\midrule
\multirow{2}{*}{Qwen3-32B}
  & w/o & 8.92 & {+3.24} & 8.02 & {+2.38} & 7.68 & {+1.68} \\
  & w/  & 8.25 & {+2.57} & 7.26 & {+1.62} & 6.79 & {+0.79} \\
\midrule
\multirow{2}{*}{Llama-3-8B}
  & w/o & 8.20 & {+2.52} & 7.66 & {+2.02} & 6.74 & {+0.74} \\
  & w/  & 7.93 & {+2.25} & 7.39 & {+1.75} & 5.94 & {$-$0.06} \\
\midrule
\multirow{2}{*}{Llama-3.1-8B}
  & w/o & 8.71 & {+3.03} & 8.03 & {+2.39} & 7.53 & {+1.53} \\
  & w/  & 8.09 & {+2.41} & 7.74 & {+2.10} & 5.79 & {$-$0.21} \\
\bottomrule
\end{tabular}
\end{table}

% In this section, we investigate the calibration of the absolute scores.
% we analyze the scoring bias of different models. Figure~\ref{fig:scoring_bias_heatmap} shows the bias of the average predicted score for each true authority score (0-9) on \domain.

% We find that models tend to overestimate the authority of low-authority sources and underestimate the authority of high-authority sources. This "regression to the mean" phenomenon confirms that although Pointwise-Judge yield reasonably strong ranking performance in the listwise setting, notable discrepancies persist with respect to their absolute prediction accuracy.

% We further conduct a case study on social media platforms (Table~\ref{tab:social_media_bias}). The results show that the significant overestimation of these site appears driven by brand recognition overriding the structural authority signal of the URL. This suggest a practical heuristic: given the prevalence of social media content and its known overestimation, setting a slightly higher filtering threshold can help counteract this specific bias.

\heading{Effective Identification of .gov/.edu Authority} 
Across all models, government and education domains consistently receive high authority scores (often in the 8-9 range), leading to notable positive divergence from their PageRank values (e.g., Qwen3-14B yields a mean divergence of +3.28 for \texttt{.gov} and +3.20 for \texttt{.edu}). 
As registration of \texttt{.gov} and \texttt{.edu} domains is strictly restricted to official government and educational entities, \texttt{.gov} and \texttt{.edu} can be regarded as inherently high-authority TLDs. The models' high scoring for such domains thus reflects accurate authority judgment. This divergence further reflects a critical limitation of PageRank as an authority proxy: it fails to capture intrinsic authority unaccompanied by a dense hyperlink structure.
For instance, a local government website may have low PageRank due to insufficient external links, yet remains the definitive authority within its domain. 

\heading{Overestimation Bias on Social Media Domains}
For social media platforms, the degree of overestimation varies substantially across models, ranging from +0.74 (Llama-3-8B, w/o text) to +2.44 (Qwen3-14B, w/o text). Notably, incorporating webpage text mitigates the overestimation of domain authority for most models: Llama-3.1-8B's bias drops from +1.53 to $-0.21$, and Llama-3-8B achieves near-zero bias ($-0.06$) with webpage text, likely because the content itself is not authoritative. 

Despite basic security measures offered by mature commercial operators, social media domains are inherently low-authority sources. As open platforms hosting user-generated content (UGC) from heterogeneous and unvetted publishers without mandatory fact-checking or official verification mechanisms, they lack institutional authority by nature. 
The overestimation without webpage text appears driven by brand recognition—well-known platforms such as \texttt{jianshu.com} and \texttt{spotify.com} (both with PageRank=0 in our dataset) receive high scores because the models associate these brand names with popularity, overriding the structural authority signal. 
This suggests a practical implication: when deploying PointJudge on social media content without access to page text, applying a slightly higher filtering threshold can help counteract this systematic overestimation.
\begin{tcolorbox}[
  breakable, enhanced,
  colback=black!2!white, colframe=black!60, boxrule=1pt, arc=2pt,
  title={\textbf{Case Study: Qwen3-32B PointJudge Scoring (w/o text)}},
  fonttitle=\bfseries\small, left=6pt, right=6pt, top=5pt, bottom=5pt
]
\small
\textbf{\textcolor{black!70}{Limitation of PageRank for Official Domains}}\\
Local government and educational sites receive low PageRank due to sparse inbound links, yet the model correctly recognizes their institutional authority:
\begin{itemize}[leftmargin=1.2em, noitemsep, topsep=2pt, parsep=0pt]
    \item \texttt{xjedu.gov.cn} (Xinjiang Edu. Dept.): PageRank\,2 $\rightarrow$ Pred\,\textbf{9} \textcolor{gray}{(+7)}
    \item \texttt{bnuz.edu.cn} (BNU Zhuhai Campus): PageRank\,3 $\rightarrow$ Pred\,\textbf{8} \textcolor{gray}{(+5)}
\end{itemize}
\smallskip
\textbf{\textcolor{black!70}{Overestimation on Social Media}}\\
Well-known social media platforms are overestimated due to brand recognition, despite being inherently low-authority open platforms:
\begin{itemize}[leftmargin=1.2em, noitemsep, topsep=2pt, parsep=0pt]
    \item \texttt{zhihu.com} (Zhihu): PageRank\,6 $\rightarrow$ Pred\,\textbf{8} \textcolor{gray}{(+2)}
    \item \texttt{tieba.com} (Baidu Tieba): PageRank\,0 $\rightarrow$ Pred\,\textbf{6} \textcolor{gray}{(+6)}
\end{itemize}
\end{tcolorbox}

\begin{table}[t]
\centering
\small
\setlength{\tabcolsep}{2.5pt}
\caption{PairJudge ranking derivation methods on \domain (fine-grained). Spearman's $\rho$ (\%). \textbf{Bold}: best; \underline{underline}: second best, per row.}
\label{tab:pairjudge_domain_fine}
\begin{tabular}{@{}llccc@{}}
\toprule
\textbf{Model} & \textbf{Ctx.} & \textbf{PairRank} & \textbf{PointScore\textsubscript{Bbl}} & \textbf{PointScore\textsubscript{Pt}} \\
\midrule
\multirow{2}{*}{Qwen3-8B}
  & w/o & 44.19 & \secondbest{64.61} & \best{71.35} \\
  & w/  & 53.05 & \secondbest{61.83} & \best{63.91} \\
\midrule
\multirow{2}{*}{Qwen3-14B}
  & w/o & \secondbest{70.21} & 69.36 & \best{73.43} \\
  & w/  & \secondbest{64.78} & 64.65 & \best{67.99} \\
\midrule
\multirow{2}{*}{Qwen3-32B}
  & w/o & 72.10 & \secondbest{72.52} & \best{75.28} \\
  & w/  & 66.32 & \secondbest{67.26} & \best{69.93} \\
\midrule
\multirow{2}{*}{Llama-3-8B}
  & w/o & \secondbest{61.05} & 59.70 & \best{64.83} \\
  & w/  & 49.06 & \secondbest{59.75} & \best{64.97} \\
\midrule
\multirow{2}{*}{Llama-3.1-8B}
  & w/o & \best{63.82} & 60.17 & \secondbest{63.24} \\
  & w/  & 51.96 & \secondbest{60.01} & \best{62.88} \\
\bottomrule
\end{tabular}
\end{table}

\begin{table}[t]
\centering
\small
\setlength{\tabcolsep}{2.5pt}
\caption{PairJudge ranking derivation methods on \domain (coarse-grained). Spearman's $\rho$ (\%). \textbf{Bold}: best; \underline{underline}: second best, per row.}
\label{tab:pairjudge_domain_coarse}
\begin{tabular}{@{}llccc@{}}
\toprule
\textbf{Model} & \textbf{Ctx.} & \textbf{PairRank} & \textbf{PointScore\textsubscript{Bbl}} & \textbf{PointScore\textsubscript{Pt}} \\
\midrule
\multirow{2}{*}{Qwen3-8B}
  & w/o & 13.34 & \secondbest{64.71} & \best{70.90} \\
  & w/  & 8.73  & \secondbest{63.12} & \best{64.84} \\
\midrule
\multirow{2}{*}{Qwen3-14B}
  & w/o & 65.65 & \secondbest{71.73} & \best{73.16} \\
  & w/  & \best{68.30} & 65.21 & \secondbest{66.55} \\
\midrule
\multirow{2}{*}{Qwen3-32B}
  & w/o & 72.88 & \secondbest{74.72} & \best{75.52} \\
  & w/  & \best{69.82} & 65.21 & \secondbest{66.55} \\
\midrule
\multirow{2}{*}{Llama-3-8B}
  & w/o & 55.09 & \secondbest{61.25} & \best{64.30} \\
  & w/  & 52.04 & \secondbest{54.92} & \best{56.07} \\
\midrule
\multirow{2}{*}{Llama-3.1-8B}
  & w/o & \secondbest{63.84} & 61.23 & \best{64.26} \\
  & w/  & 57.08 & \secondbest{58.12} & \best{58.69} \\
\bottomrule
\end{tabular}
\end{table}

\begin{table}[t]
\centering
\small
\setlength{\tabcolsep}{1.8pt}
\caption{PairJudge ranking derivation methods on \entity (fine-grained). Spearman's $\rho$ (\%). \textbf{Bold}: best; \underline{underline}: second best, per row.}
\label{tab:pairjudge_entity_fine}
\begin{tabular}{@{}llccc@{}}
\toprule
\textbf{Domain} & \textbf{Model} & \textbf{PairRank} & \textbf{PointScore\textsubscript{Bbl}} & \textbf{PointScore\textsubscript{Pt}} \\
\midrule
\multirow{5}{*}{\rotatebox[origin=c]{90}{\textbf{Basketball}}}
  & Qwen3-8B     & 81.29 & \secondbest{83.28} & \best{86.67} \\
  & Qwen3-14B    & \secondbest{85.09} & 82.16 & \best{85.89} \\
  & Qwen3-32B    & 82.25 & \secondbest{82.77} & \best{85.94} \\
  & Llama-3-8B   & 83.54 & \secondbest{87.14} & \best{88.90} \\
  & Llama-3.1-8B & 79.29 & \secondbest{84.59} & \best{87.43} \\
\midrule
\multirow{5}{*}{\rotatebox[origin=c]{90}{\textbf{Movies}}}
  & Qwen3-8B     & 61.33 & \secondbest{81.51} & \best{87.87} \\
  & Qwen3-14B    & \secondbest{86.19} & 86.45 & \best{88.60} \\
  & Qwen3-32B    & 83.27 & \secondbest{87.02} & \best{89.56} \\
  & Llama-3-8B   & 79.04 & \secondbest{82.19} & \best{83.24} \\
  & Llama-3.1-8B & 70.00 & \secondbest{81.00} & \best{82.72} \\
\midrule
\multirow{5}{*}{\rotatebox[origin=c]{90}{\textbf{Songs}}}
  & Qwen3-8B     & 10.57 & \secondbest{37.06} & \best{44.58} \\
  & Qwen3-14B    & 47.62 & \secondbest{48.04} & \best{50.54} \\
  & Qwen3-32B    & 39.92 & \secondbest{49.88} & \best{51.66} \\
  & Llama-3-8B   & 25.62 & \secondbest{46.49} & \best{48.33} \\
  & Llama-3.1-8B & 40.27 & \secondbest{45.87} & \best{46.76} \\
\bottomrule
\end{tabular}
\end{table}

\begin{table}[t]
\centering
\small
\setlength{\tabcolsep}{1.8pt}
\caption{PairJudge ranking derivation methods on \entity (coarse-grained). Spearman's $\rho$ (\%). \textbf{Bold}: best; \underline{underline}: second best, per row.}
\label{tab:pairjudge_entity_coarse}
\begin{tabular}{@{}llccc@{}}
\toprule
\textbf{Domain} & \textbf{Model} & \textbf{PairRank} & \textbf{PointScore\textsubscript{Bbl}} & \textbf{PointScore\textsubscript{Pt}} \\
\midrule
\multirow{5}{*}{\rotatebox[origin=c]{90}{\textbf{Basketball}}}
  & Qwen3-8B     & 29.52 & \secondbest{83.50} & \best{84.19} \\
  & Qwen3-14B    & \best{86.50} & 82.84 & \secondbest{83.80} \\
  & Qwen3-32B    & \best{84.76} & \secondbest{84.21} & 83.94 \\
  & Llama-3-8B   & 83.91 & \secondbest{86.92} & \best{87.46} \\
  & Llama-3.1-8B & 78.53 & \secondbest{85.83} & \best{86.52} \\
\midrule
\multirow{5}{*}{\rotatebox[origin=c]{90}{\textbf{Movies}}}
  & Qwen3-8B     & 48.32 & \secondbest{65.22} & \best{86.75} \\
  & Qwen3-14B    & \best{89.03} & \secondbest{88.66} & 88.77 \\
  & Qwen3-32B    & \secondbest{85.90} & 90.15 & \best{90.27} \\
  & Llama-3-8B   & 80.84 & \secondbest{87.04} & \best{87.42} \\
  & Llama-3.1-8B & 67.35 & \secondbest{77.78} & \best{80.84} \\
\midrule
\multirow{5}{*}{\rotatebox[origin=c]{90}{\textbf{Songs}}}
  & Qwen3-8B     & 24.58 & \best{41.01} & \secondbest{39.27} \\
  & Qwen3-14B    & 48.89 & \secondbest{49.14} & \best{50.77} \\
  & Qwen3-32B    & 39.60 & \secondbest{51.22} & \best{51.42} \\
  & Llama-3-8B   & 40.20 & \secondbest{49.58} & \best{49.65} \\
  & Llama-3.1-8B & 19.50 & \secondbest{46.32} & \best{48.41} \\
\bottomrule
\end{tabular}
\end{table}

\section{Comparison of PairJudge Ranking Derivation Methods}
\label{app:pairjudge_comparison}

To validate our choice of ranking derivation method for \texttt{PairJudge}, we compare three distinct approaches for converting pairwise judgments into a listwise ranking:
\begin{itemize}[leftmargin=*,itemsep=2pt,topsep=3pt,parsep=0pt]
    \item \textbf{PairRank (Bubble Sort)}: Applies bubble sort directly to the binary outcomes (A > B or B > A) of pairwise comparisons.
    \item \textbf{PointScore (Bubble Sort)}: First elicits a score for each item in a pair (e.g., A:7, B:4), then applies bubble sort based on which item received the higher score in each comparison.
    \item \textbf{PointScore (Point-based)}: Ranks items based on their average score received across all comparisons. This is the method we refer to as \textbf{PointScore} in the main paper.
\end{itemize}

Tables~\ref{tab:pairjudge_domain_fine}, \ref{tab:pairjudge_domain_coarse},~\ref{tab:pairjudge_entity_fine} and~\ref{tab:pairjudge_entity_coarse} present the results across all datasets. The point-based \textbf{PointScore} method consistently outperforms the other two, especially the \textbf{PairRank} bubble sort. This is because averaging scores is more robust to the transitivity errors that can arise from isolated, inconsistent pairwise judgments, thus providing a more stable and reliable final ranking.

\end{document}